\documentclass[aps,prc,twocolumn]{revtex4-1}

\usepackage{wallpaper}
\usepackage[colorlinks,linkcolor=blue,anchorcolor=blue,citecolor=red]{hyperref}
\usepackage{amsmath}
\usepackage{multirow}

\begin{document}
\hyphenpenalty=6000
\tolerance=1000

\hyphenation{Eq}

\title{Nuclear pasta structures at high temperatures}
\author{Cheng-Jun~Xia$^{1}$}
\email{cjxia@yzu.edu.cn}
\author{Toshiki Maruyama$^{2}$}
\email{maruyama.toshiki@jaea.go.jp}
\author{Nobutoshi Yasutake$^{3, 2}$}
\email{nobutoshi.yasutake@it-chiba.ac.jp}
\author{Toshitaka Tatsumi$^{4}$}
\email{tatsumitoshitaka@gmail.com}

\affiliation{$^{1}${Center for Gravitation and Cosmology, College of Physical Science and Technology, Yangzhou University, Yangzhou 225009, China}
\\$^{2}${Advanced Science Research Center, Japan Atomic Energy Agency, Shirakata 2-4, Tokai, Ibaraki 319-1195, Japan}
\\$^{3}${Department of Physics, Chiba Institute of Technology (CIT), 2-1-1 Shibazono, Narashino, Chiba, 275-0023, Japan}
\\$^{4}${Institute of Education, Osaka Sangyo University, 3-1-1 Nakagaito, Daito, Osaka 574-8530, Japan}}

\date{\today}

\begin{abstract}
We investigate nuclear pasta structures at high temperatures in the framework of relativistic mean field model with Thomas-Fermi approximation. Typical pasta structures (droplet, rod, slab, tube, and bubble) are obtained, which form various crystalline configurations. The properties of those nuclear pastas are examined in a three-dimensional geometry with reflection symmetry, where the optimum lattice constants are fixed by reproducing the droplet/bubble density that minimizes the free energy adopting spherical or cylindrical approximations for Wigner-Seitz cells.  It is found that different crystalline structures can evolve into each other via volume conserving deformations. For fixed densities and temperatures, the differences of the free energies per baryon of nuclear pasta in various shapes and lattice structures are typically on the order of tens of keV, suggesting the possible coexistence of those structures. As temperature increases, the thermodynamic fluctuations are expected to disrupt the long-range ordering in nuclear pasta structures. We then estimate the critical conditions for nuclear pasta to become disordered and behave like liquid, which are found to be sensitive to the densities, temperatures, proton fractions, and nuclear shapes. If we further increase temperature, eventually the nonuniform structures of nuclear pasta become unstable and are converted into uniform nuclear matter. The phase diagrams of nuclear matter are then estimated, which should be useful for understanding the evolutions of neutron stars, supernova dynamics, and binary neutron star mergers.
\end{abstract}

\maketitle

\section{\label{sec:intro}Introduction}
Due to the extreme pressure resides in neutron stars and supernovae at the stage of gravitational collapse, the stellar matter is compressed so extensively that nuclei come into close contact. At such large densities, neutrons may drip out of nuclei and form neutron gas, then the dense stellar matter is  essentially a liquid-gas mixed phase of nuclear matter, which typically forms a lattice of spherical nuclei emersed in a gas of electrons and neutrons. If we further increase the density, the Coulomb repulsion between nuclei becomes so intense that it is energetically favorable for nuclei to be deformed, which form various exotic shapes that resemble pasta, i.e., nuclear pasta~\cite{Baym1971_ApJ170-299, Negele1973_NPA207-298, Ravenhall1983_PRL50-2066, Hashimoto1984_PTP71-320, Williams1985_NPA435-844}. It was shown that the nuclear pasta should exist up to the densities $n_\mathrm{b}\lesssim 0.1\ \mathrm{fm}^{-3}$ and temperatures $T\lesssim 15$ MeV, beyond which the uniform phases are more stable~\cite{Yang2019_PRC100-054314, Yang2021_PRC103-014304, Shen2011_ApJ197-20, Togashi2017_NPA961-78}.

The microscopic structures of the pasta phase play important roles in the transport and elastic properties of dense stellar matter, which may be identified in various astrophysical scenarios~\cite{Chamel2008_LRR11-10, Caplan2017_RMP89-041002}. For example, the neutrino-pasta scattering affects the neutrino opacity~\cite{Horowitz2004_PRC69-045804, Schuetrumpf2020_PRC101-055804}, which may lead to late-time enhancement of the neutrino luminosity in core-collapse supernovae~\cite{Horowitz2016} and alter the cooling processes in neutron stars~\cite{Gusakov2004_AA421-1143, Carvalho2014_PRC90-055804}. The complex shapes of nuclear pasta act like impurities for electron scattering~\cite{Schneider2016_PRC93-065806}, which effectively dissipates the electric currents that support the magnetic fields in neutron stars~\cite{Pons2013_NP9-431, Gao2017_ApJ849-19}. According to the Wiedemann-Franz law, the thermal conductivity is linearly related to the electrical conductivity, so that nuclear pasta acts as a thermally resistive layer inside a neutron star, leading to longer cooling time in quiescent low mass x-ray binaries~\cite{Horowitz2015_PRL114-031102}. The interplay of nuclear pasta with the superfluid vortexes was essential to explain pulsar glitches in the framework of vortex creep model, which affects the glitch sizes as well as the post glitch recovery behaviors~\cite{Anderson1975_Nature256-25, Lorenz1993_PRL70-379, Mochizuki1995_ApJ440-263, Horowitz2004_PRC69-045804, Gusakov2004_AA421-1143, Gearheart2011_MNRAS418-2343, Rueda2014_PRC89-035804, Carvalho2014_PRC90-055804, Wlazlowski2016_PRL117-232701, Watanabe2017_PRL119-062701, Sekizawa2022_PRC105-045807}. A crust failure could trigger the sudden release of magnetic and elastic energy observed in magnetar bursts~\cite{Beloborodov2014_ApJ794-L24, Beloborodov2016_ApJ833-261, Li2016_ApJ833-189}, the short gamma-ray burst precursors of neutron star mergers~\cite{Tsang2012_PRL108-011102}, as well as pulsar glitches~\cite{Ruderm1969_Nature223-597, Baym1971_AP66-816, Haskell2015_IJMPD24-1530008, Akbal2017_MNRAS473-621, Guegercinoglu2019_MNRAS488-2275, Layek2020_MNRAS499-455}. The quasi-periodic oscillations observed after giant flares of soft gamma repeaters are usually interpreted as global oscillations of magnetars~\cite{Kouveliotou1998_Nature393-235, Hurley1999_Nature397-41}. It was shown that the oscillation spectrum is affected by the elastic and superfluid properties of nuclear pasta~\cite{Hansen1980_ApJ238-740, Schumaker1983_MNRAS203-457, McDermott1988_ApJ325-725,  Strohmayer1991_ApJ375-679, Passamonti2012_MNRAS419-638, Gabler2018_MNRAS476-4199, Sotani2012_PRL108-201101, Sotani2016_MNRAS464-3101, Sotani2019_MNRAS489-3022, Pethick2020_PRC101-055802, Kozhberov2020_MNRAS498-5149}. Due to the elastic stresses of astromaterials, there could be mountains on neutron stars, which are responsible for the asymmetric matter distributions with nonzero ellipticities $\epsilon$. The sizes of mountains are limited by the breaking strain of neutron star crust, which can reach as large as 0.1 for nuclear pasta~\cite{Horowitz2009_PRL102-191102, Chugunov2010_MNRAS407-L54, Horowitz2010_PRD81-103001, Caplan2018_PRL121-132701, Baiko2018_MNRAS480-5511, Kozhberov2020_MNRAS498-5149}. In such cases, it is possible that the maximum ellipticity of neutron stars reaches $\epsilon\approx 10^{-5}$~\cite{Ushomirsky2000_MNRAS319-902}, which are expected to emit gravitational waves for fast rotating neutron stars. In fact, recent observations have put strong constraints on the maximum amplitude of such gravitational waves, where the ellipticities for the recycled pulsars PSR J0437-4715 and PSR J0711-6830 are constrained with $\epsilon < 10^{-8}$~\cite{Abbott2020_ApJ902-L21}, indicating the absence of large mountains.

Adopting spherical or cylindrical approximations of the Wigner-Seitz (WS) cell~\cite{Pethick1998_PLB427-7, Oyamatsu1993_NPA561-431, Maruyama2005_PRC72-015802, Avancini2008_PRC78-015802, Avancini2009_PRC79-035804, Grill2012_PRC85-055808, Gupta2013_PRC87-028801, Togashi2017_NPA961-78, Shen2011_ApJ197-20, Xia2022_PRC105-045803, Xia2022_CTP74-095303, Parmar2022_PRD106-023031}, it was found that nuclear pasta exhibits at least five types of geometrical structures, i.e, droplets, rods, slabs, tubes, and bubbles. Considering the interactions among cells, those exotic nuclei arrange themselves into various types of lattice structures, where the body-centered cubic (BCC) lattices for droplets/bubbles and the honeycomb lattices for rods/tubes were found to be more stable~\cite{Oyamatsu1984_PTP72-373}. Carrying out more detailed investigations on nuclear pasta structures in a three-dimensional geometry, it was shown that the face-centered cubic (FCC) lattices could become energetically favorable for droplets/bubbles, which were obtained in a unified manner in the framework of Thomas-Fermi approximation~\cite{Okamoto2012_PLB713-284, Okamoto2013_PRC88-025801, Xia2021_PRC103-055812}. Additionally, the intermediate structures of droplets and rods, slabs and tubes were identified during the transition between those phases~\cite{Watanabe2003_PRC68-035806, Okamoto2012_PLB713-284}. Much more complicated structures were observed as well~\cite{Magierski2002_PRC65-045804, Newton2009_PRC79-055801, Fattoyev2017_PRC95-055804}, e.g., the gyroid and double-diamond morphologies~\cite{Nakazato2009_PRL103-132501, Schuetrumpf2015_PRC91-025801}, P-surface configurations~\cite{Schuetrumpf2013_PRC87-055805, Schuetrumpf2019_PRC100-045806}, nuclear waffles~\cite{Schneider2014_PRC90-055805, Sagert2016_PRC93-055801}, Parking-garage structures~\cite{Berry2016_PRC94-055801}, and deformations in droplets~\cite{Kashiwaba2020_PRC101-045804}.

\begin{table}
\caption{\label{tab:NM} Temperatures and densities at the critical end points for the liquid-gas transition of nuclear matter at various proton fractions $Y_p$, which are predicted by the covariant density functional MTVTC~\cite{Maruyama2005_PRC72-015802}. }
\begin{tabular}{c|ccc} \hline \hline
$Y_p$                          &   0.1     &    0.3    &   0.5  \\ \hline
$n_\mathrm{b}$ (fm${}^{-3}$)   &  0.027    &   0.042   &  0.047 \\
$T$ (MeV)                      &  6.45     &   13.5    &  16.1 \\
\hline
\end{tabular}
\end{table}

In this work we further investigate the effects of finite temperature on nuclear pasta structures, where the Thomas-Fermi approximation is adopted. The local properties of nuclear matter are fixed within the framework of the relativistic mean field (RMF) model~\cite{Meng2016_RDFNS}, where the covariant density functional MTVTC is adopted~\cite{Maruyama2005_PRC72-015802}. This functional predicts the symmetry energy $S = 32.46$ MeV and its slope $L = 89.39$ MeV of nuclear matter at saturation density $n_0= 0.153$ fm${}^{-3}$, while the corresponding critical temperatures and densities of the liquid-gas phase transition are indicated in Table~\ref{tab:NM}. Note that the charge number of nuclei, the core-crust transition density, and the onset density of non-spherical nuclei decrease with $L$~\cite{Oyamatsu2007_PRC75-015801, Ducoin2011_PRC83-045810, Ji2020_PRC102-015806}, so that the results obtained in this work may be altered if we decrease $L$ explicitly by introducing an $\omega$-$\rho$ cross coupling term~\cite{Xia2021_PRC103-055812}. According to our previous investigations~\cite{Xia2021_PRC103-055812}, the effect of finite cell size~\cite{GimenezMolinelli2014_NPA923-31, Newton2009_PRC79-055801} needs to be considered by searching for the optimum size of a unit cell, where the corresponding volume occupied by each droplet/bubble was found to take the same value in disregard of the exact lattice structures. In such cases, we first search for the optimum WS cell size adopting spherical or cylindrical approximations, then fix the optimum size of the unit cell according to the lattice structure, which significantly reduces the numerical cost. The paper is organized as follows. In Sec.~\ref{sec:the}, we present our theoretical framework. The obtained results on the structure and properties of nuclear pasta are presented in Sec.~\ref{sec:pasta}. Our conclusion is given in Sec.~\ref{sec:con}.

\section{\label{sec:the} Theoretical framework}
We adopt the following Lagrangian density of RMF model for the investigation of hot nuclear matter, i.e.,
\begin{eqnarray}
\mathcal{L}
 &=& \sum_{i=n,p} \bar{\psi}_i
       \left[  i \gamma^\mu \partial_\mu - \gamma^0 \left(g_\omega\omega + g_\rho\rho\tau_i + A q_i\right)- M^* \right] \psi_i
\nonumber \\
 &&\mbox{} + \bar{\psi}_e \left[ i \gamma^\mu \partial_\mu - m_e - q_e \gamma^0 A \right]\psi_e - \frac{1}{4} A_{\mu\nu}A^{\mu\nu}
\nonumber \\
 &&\mbox{} + \frac{1}{2}\partial_\mu \sigma \partial^\mu \sigma  - \frac{1}{2}m_\sigma^2 \sigma^2
           - \frac{b}{3} M (g_{\sigma} \sigma)^3 - \frac{c}{4} (g_{\sigma} \sigma)^4
\nonumber \\
 &&\mbox{} - \frac{1}{4} \omega_{\mu\nu}\omega^{\mu\nu} + \frac{1}{2}m_\omega^2 \omega^2
           - \frac{1}{4} \rho_{\mu\nu}\rho^{\mu\nu} + \frac{1}{2}m_\rho^2 \rho^2,
\label{eq:Lagrange}
\end{eqnarray}
where $\tau_n=-\tau_p=1$ is the 3rd component of isospin, $q_p = - q_e = e$ and $q_n=0$ the charge, and $M^*\equiv M + g_{\sigma} \sigma$ the effective nucleon mass. In this work we adopt the covariant density functional MTVTC, where the coefficients in Eq.~(\ref{eq:Lagrange}) can be found in Ref.~\cite{Maruyama2005_PRC72-015802}. The boson fields $\sigma$, $\omega$, $\rho$, and $A$ take mean values with only the time components due to time-reversal symmetry. Then the field tensors $\omega_{\mu\nu}$, $\rho_{\mu\nu}$, and $A_{\mu\nu}$ vanish except for
\begin{equation}
\omega_{i0} = -\omega_{0i} = \partial_i \omega,
 \rho_{i0}  = -\rho_{0i}   = \partial_i  \rho,
  A_{i0}    = -A_{0i}      = \partial_i A.\nonumber
\end{equation}
Based on the Euler-Lagrange equation, the Klein-Gordon equations for bosons in the framework of mean field approximation are
\begin{eqnarray}
(-\nabla^2 + m_\sigma^2) \sigma &=& - g_{\sigma} \left(n_\mathrm{b}^s + b M g_{\sigma} \sigma^2 + c  g_{\sigma}^2 \sigma^3  \right), \label{eq:KG_sigma} \\
(-\nabla^2 + m_\omega^2) \omega &=& g_{\omega} n_\mathrm{b}, \label{eq:KG_omega}\\
(-\nabla^2 + m_\rho^2) \rho &=&  g_{\rho} \left(n_n-n_p\right), \label{eq:KG_rho}\\
- \nabla^2 A &=& e n_p - e n_e. \label{eq:KG_photon}
\end{eqnarray}
Here $n_\mathrm{b}^s = n^s_p+n^s_n$ and $n_\mathrm{b} = n_p+n_n$ are the local scalar and vector densities of nucleons in the Thomas-Fermi approximation, which are fixed by
\begin{eqnarray}
n^s_i &=& \frac{m^*_i}{\pi^2}\int_0^\infty   \left[f_i^+(p)+f_i^-(p)\right] \frac{p^2}{\sqrt{p^2 + {m_i^*}^2}} \mbox{d}p,\\
n_i  &=& \frac{1}{\pi^2}\int_0^\infty \left[f_i^+(p) - f_i^-(p) \right] p^2 \mbox{d}p.
\end{eqnarray}
Note that we have adopted the Fermi-Dirac distribution for nucleons and electrons with
\begin{equation}
  f_i^\pm(p) = \left[1 + {\rm e}^{(\sqrt{p^2 + {m_i^*}^2}\mp\mu_i^*)/T} \right]^{-1},
\end{equation}
where $\mu_i^*$ represents the effective chemical potential and $T$ the temperature. The real chemical potential can then be obtained by including the vector potentials, i.e.,
\begin{equation}
\mu_i(\vec{r}) = \mu_i^*(\vec{r})  + g_{\omega} \omega + g_{\rho}\tau_{i} \rho + q_i  A. \label{eq:chem}
\end{equation}
Note that if the interactions among particles are absent, we have a free system and the effective chemical potential is equivalent to the real one, i.e., $\mu_i = \mu_i^*$. The total particle number, entropy, and free energy of the system are determined by
\begin{equation}
N_i=\int n_i(\vec{r}) \mbox{d}^3 r, \ S=\sum_i\int s_i(\vec{r}) \mbox{d}^3 r,\  F=\int {\cal{F}}(\vec{r}) \mbox{d}^3 r,   \label{eq:energy}
\end{equation}
with
\begin{eqnarray}
\cal{F}
&=& \sum_{i=n,p,e} (\varepsilon_i - T s_i)+ \frac{1}{2}(\nabla \sigma)^2 + \frac{1}{2}m_\sigma^2 \sigma^2    \nonumber \\
&&    +\frac{b}{3} M (g_{\sigma} \sigma)^3 + \frac{c}{4} (g_{\sigma} \sigma)^4 + \frac{1}{2}(\nabla \omega)^2 + \frac{1}{2}m_\omega^2 \omega^2  \nonumber \\
&&    + \frac{1}{2}(\nabla \rho)^2 + \frac{1}{2}m_\rho^2 \rho^2 + \frac{1}{2}(\nabla A)^2, \label{eq:ener_dens}\\
\varepsilon_i &=& \frac{1}{\pi^2}\int_0^\infty   \left[f_i^+(p)+f_i^-(p)\right] \sqrt{p^2 + {m_i^*}^2} p^2 \mbox{d}p,\\
s_i &=& \frac{1}{\pi^2}\int_0^\infty \left\{ f_i^+(p)\ln\left[\frac{1}{f_i^+(p)} - 1\right] - \ln\left[f_i^+(p)f_i^-(p)\right]    \right.\nonumber\\
     &&  \left.  + f_i^-(p)\ln\left[\frac{1}{f_i^-(p)} - 1\right]   - \frac{2}{T}\sqrt{p^2+{m_i^*}^2}    \right\} p^2 \mbox{d}p,   \label{eq:entropy}
\end{eqnarray}
where $m_e^*=m_e = 0.511$ MeV and $m_n^*=m_p^*=M^*$. The energy and pressure can then be obtained with
\begin{eqnarray}
  E &=& F+TS, \\
  P &=& \left(\sum_i\mu_iN_i-F\right)/V.
\end{eqnarray}

For any density profiles, the Klein-Gordon equations~(\ref{eq:KG_sigma}-\ref{eq:KG_photon}) can be solved via fast cosine transformations, which fulfill the reflective boundary conditions. By minimizing the total free energy $F$ at given total particle numbers $N_i$, temperature $T$, and cell size, it is found that the density distributions of fermions $n_i(\vec{r})$ follow the constancy of chemical potentials with
\begin{equation}
  \mu_i(\vec{r})= \mathrm{constant}. \label{eq:chem_cons}
\end{equation}
Note that the density profiles for protons and electrons do not follow the local charge neutrality condition $n_p(\vec{r})=n_e(\vec{r})$ for Eq.~(\ref{eq:KG_photon}). Instead, the quasineutrality condition is always fulfilled with $\sum_iq_iN_i=0$, leading to $N_p=N_e$ for the unit cells and WS cells considered here. In order to fulfill Eq.~(\ref{eq:chem_cons}), we readjust the density profiles via imaginary time step method~\cite{Levit1984_PLB139-147} and solve Eqs.~(\ref{eq:KG_sigma}-\ref{eq:KG_photon}), (\ref{eq:chem}), and~(\ref{eq:chem_cons}) iteratively. The iteration stops until the deviation of local chemical potentials from  Eq.~(\ref{eq:chem_cons}) becomes insignificant. In practice, we first fix the optimum WS cell size $R_\mathrm{W}$ adopting spherical or cylindrical approximations for WS cells~\cite{Xia2022_PRC105-045803, Xia2022_CTP74-095303}, which minimizes the free energy per nucleon at fixed average baryon number density $n_\mathrm{b}$, temperature $T$, and proton fraction $Y_p$. Then we consider the simple cubic (SC), BCC, and FCC lattices for droplets or bubbles, simple and honeycomb configurations for rods or tubes, and slabs in a three-dimensional geometry, where the optimum unit cell sizes are fixed by
\begin{eqnarray}
 && \text{3D: } 3a^3 =
 \left\{\begin{array}{l}
   4\pi R_\mathrm{W}^3, \ \ \ \  \text{SC}\\
   8\pi R_\mathrm{W}^3, \ \ \   \text{BCC}\\
   16\pi R_\mathrm{W}^3, \ \  \text{FCC}\\
 \end{array}\right.; \label{Eq:a_3D} \\
 &&\text{2D: } a^2 =
 \left\{\begin{array}{l}
   \pi R_\mathrm{W}^2,  \ \ \ \ \ \ \ \ \ \text{Simple}\\
   2\pi R_\mathrm{W}^2/\sqrt{3}, \ \  \text{Honeycomb}\\
 \end{array}\right.; \label{Eq:a_2D} \\
 &&\text{1D: } a = 2 R_\mathrm{W}, \ \ \ \  \ \ \ \text{Slab}. \label{Eq:a_1D}
\end{eqnarray}
Here $a$ represents the lattice constant of cubic unit cells while for honeycomb configurations a cuboid unit cell is adopted with $b=\sqrt{3}a$. Note that a slight deviation of the optimum lattice constant $a$ from that in Eqs.~(\ref{Eq:a_3D}-\ref{Eq:a_1D}) may persists, which is nonetheless insignificant due to the rather small energy differences among various types of lattice structures~\cite{Oyamatsu1984_PTP72-373, Xia2021_PRC103-055812}. More detailed discussions on the numerical recipe in obtaining various nuclear pasta structures can be found in our previous studies~\cite{Xia2021_PRC103-055812, Xia2022_PRC105-045803, Xia2022_CTP74-095303}.

\section{\label{sec:pasta} Results and Discussion}

\subsection{\label{sec:pasta_bulk} Shapes of nuclei}

\begin{figure}
\centering
\includegraphics[width=0.8\linewidth]{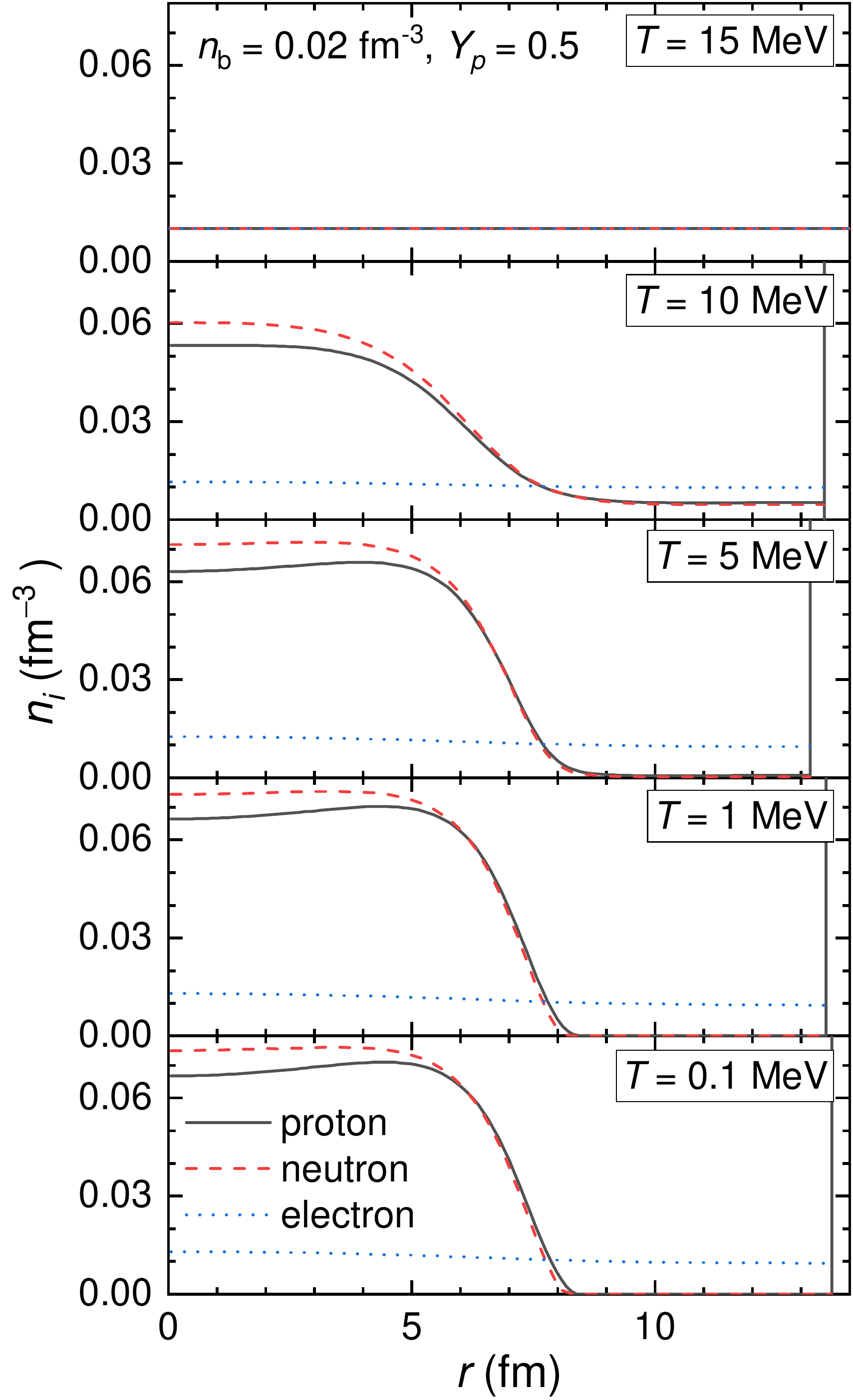}
\caption{\label{Fig:Dens} Density profiles of nucleons and electrons in WS cells of droplet phases at $T=0.1$, 1, 5, 10, 15 MeV (from bottom to top), respectively. The boundary of WS cell at $r=R_\mathrm{W}$ is indicated by a vertical line in each panel.}
\end{figure}

We first investigate the properties of nuclear pasta adopting spherical or cylindrical approximations for WS cells, where symmetric nuclear matter with proton fraction $Y_p =0.5$ and asymmetric nuclear matter with $Y_p =0.3$ and 0.1 are considered. The effects of finite temperature are examined by taking $T=0.1$, 1, 5, 10, and 15 MeV, which destabilize the nonuniform structures as $T$ increases. To show this explicitly, as an example, in Fig.~\ref{Fig:Dens} we present the density profiles of nuclear droplet phases at $n_\mathrm{b}= 0.02\ \mathrm{fm}^{-3}$, $Y_p =0.5$, and various $T$, where the WS cells are assumed to be spherical. The center of the spherical droplets is locate at $r=0$, while the cell boundary at $r=R_\mathrm{W}$ is indicated by a vertical line. The effects of charge screening are considered with nonuniform electron density distributions inside WS cells, which affects the properties of nuclear pasta~\cite{Maruyama2005_PRC72-015802}. Note that at large densities and proton fractions, the chemical potential of electrons will surpass the mass of muons, which inevitably leads to the creation of muons and contributes to the charge screening effects. The contributions of muons will be considered in our future study, while for now only electrons are considered. As temperature increases, the density of nucleons at $r=0$ decreases. At $T\gtrsim 1$ MeV, a gas comprised of protons and neutrons is formed outside of the droplet with its density increases with $T$. Consequently, the density profiles in the surface regions of nuclei become smooth, e.g., at $T=10$ MeV, which is expected to reduce the surface tension between the liquid and gas phases of nuclear matter~\cite{Maruyama2010_NPA834-561c}. If we further increase the temperature to $T=15$ MeV, the droplet phase becomes unstable and is converted into the uniform phase of nuclear matter. Meanwhile, we note that the optimum WS cell size $R_\mathrm{W}$ remains almost constant for different $T$, which would increase drastically if $T$ approaches to the uniform-nonuniform phase boundaries as illustrated in Fig.~\ref{Fig:Mic}.

\begin{figure*}[htbp]
\begin{minipage}[t]{0.327\linewidth}
\centering
\includegraphics[width=\textwidth]{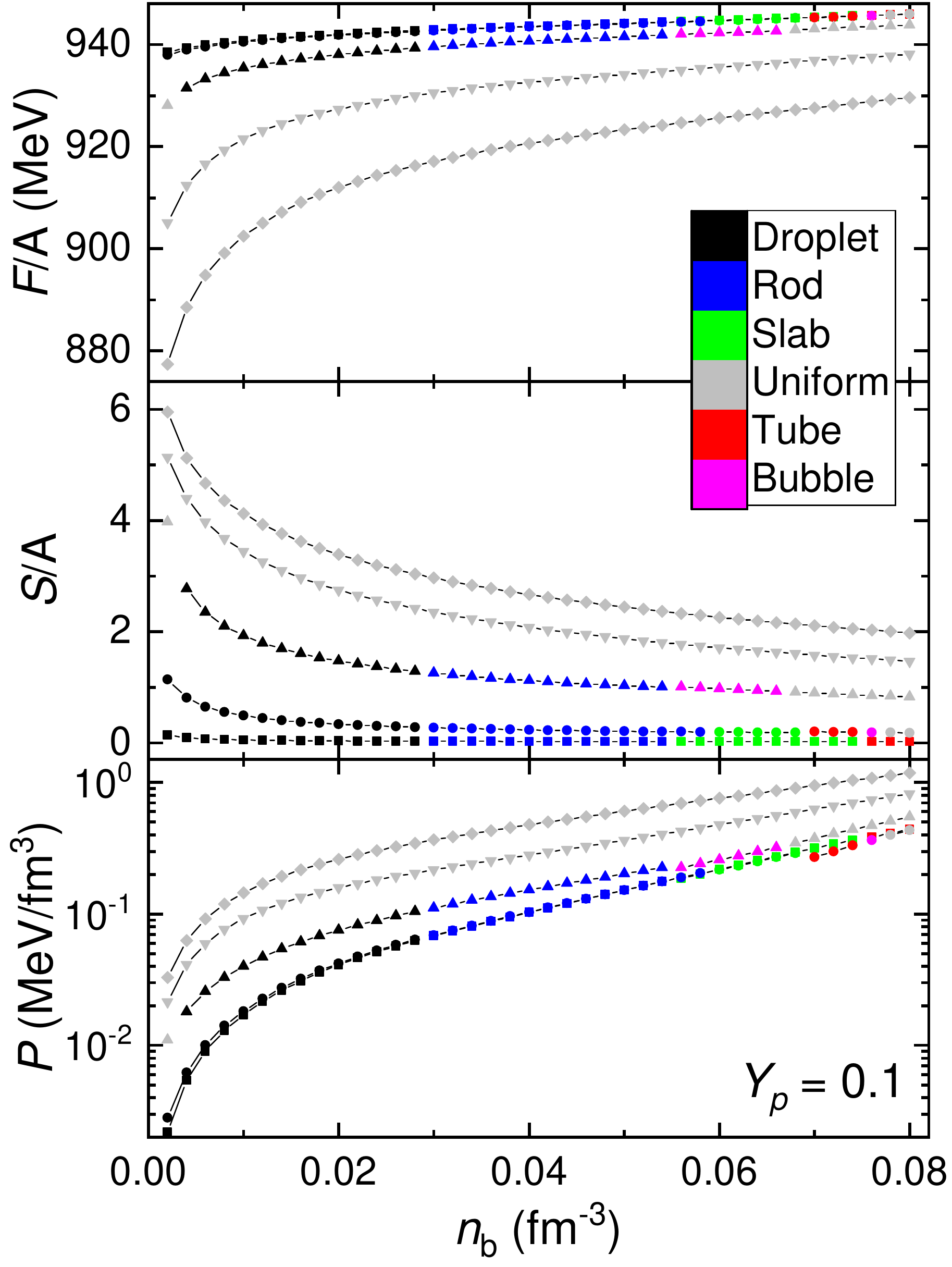}
\end{minipage}%
\hfill
\begin{minipage}[t]{0.32\linewidth}
\centering
\includegraphics[width=\textwidth]{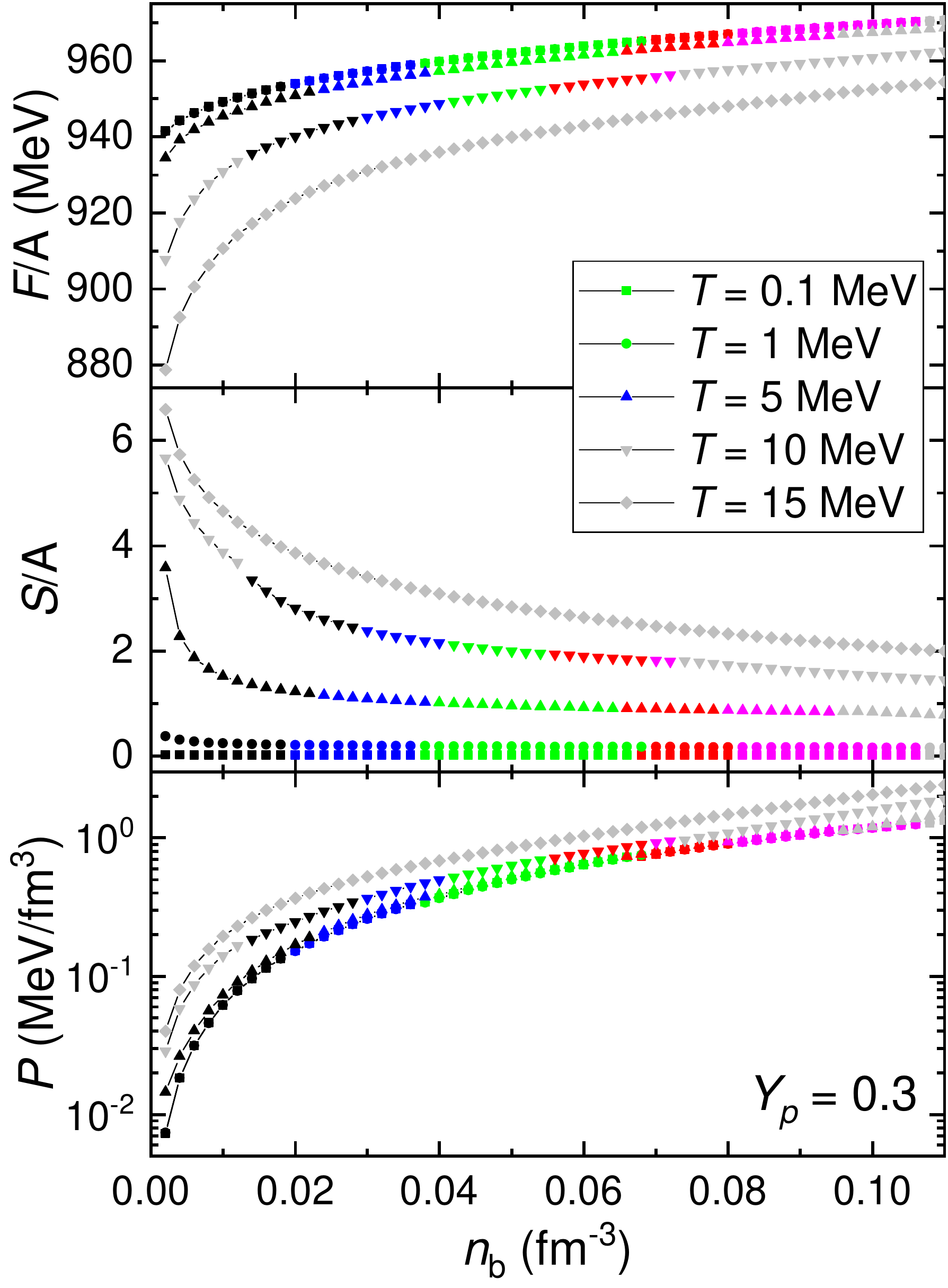}
\end{minipage}
\hfill
\begin{minipage}[t]{0.333\linewidth}
\centering
\includegraphics[width=\textwidth]{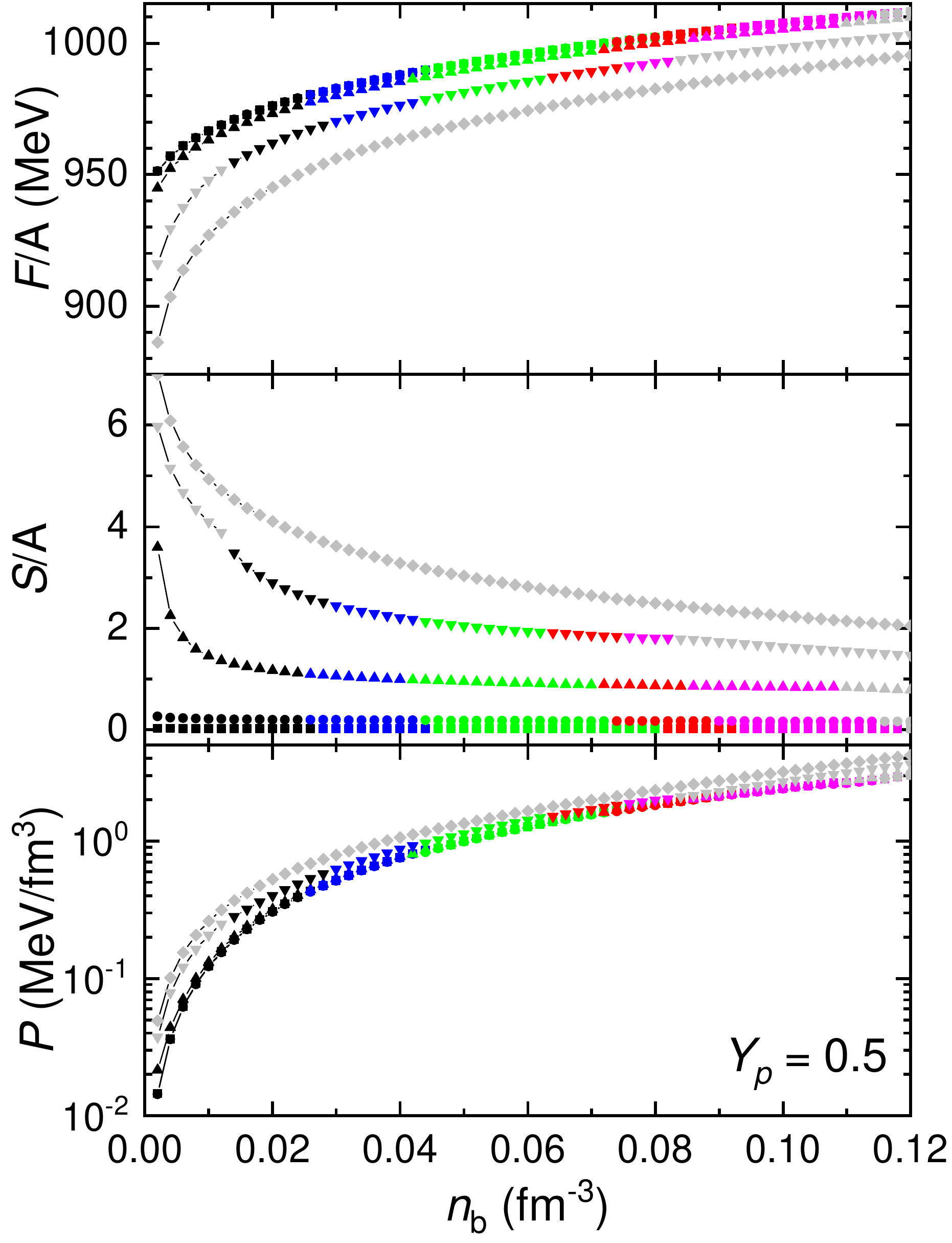}
\end{minipage}
\caption{\label{Fig:FpA} Free energy per baryon, entropy per baryon, and pressure for nuclear matter obtained with fixed proton fractions $Y_p =0.1$, 0.3, and 0.5. The types of nuclear matter structures are indicated with colors, while temperature is marked with different symbol shapes.}
\end{figure*}

Once the optimum configuration of nuclear pasta at given $n_\mathrm{b}$, $Y_p$, and $T$ is fixed, the free energy, entropy, energy, and pressure can be determined by Eqs.~(\ref{eq:chem}-\ref{eq:entropy}). In Fig.~\ref{Fig:FpA}, we present the obtained free energy per baryon, entropy per baryon, and pressure for nuclear matter in optimum configurations, where the free energy per baryon are minimized with respect to the pasta structures and WS cell sizes $R_\mathrm{W}$. In general, the free energy per baryon and pressure increase with baryon number density $n_\mathrm{b}$, while the entropy per baryon decreases. If nonuniform phases emerge for nuclear matter, the droplet, rod, slab, tube, and bubble phases appear sequentially as density increases. The density range of those nonuniform phases increases with proton fraction $Y_p$, while the free energy per baryon, entropy per baryon, and pressure increases as well. As temperature $T$ increases, the energy per baryon, entropy per baryon, and pressure of nuclear matter increase, while the free energy per baryon decreases. As illustrated in Fig.~\ref{Fig:Dens}, increasing $T$ will destablize the nonuniform structures of nuclear matter. Consequently, the density range of nonuniform nuclear matter decreases with $T$ and vanishes at $T\gtrsim 15$ MeV.

\begin{figure}
\centering
\includegraphics[width=\linewidth]{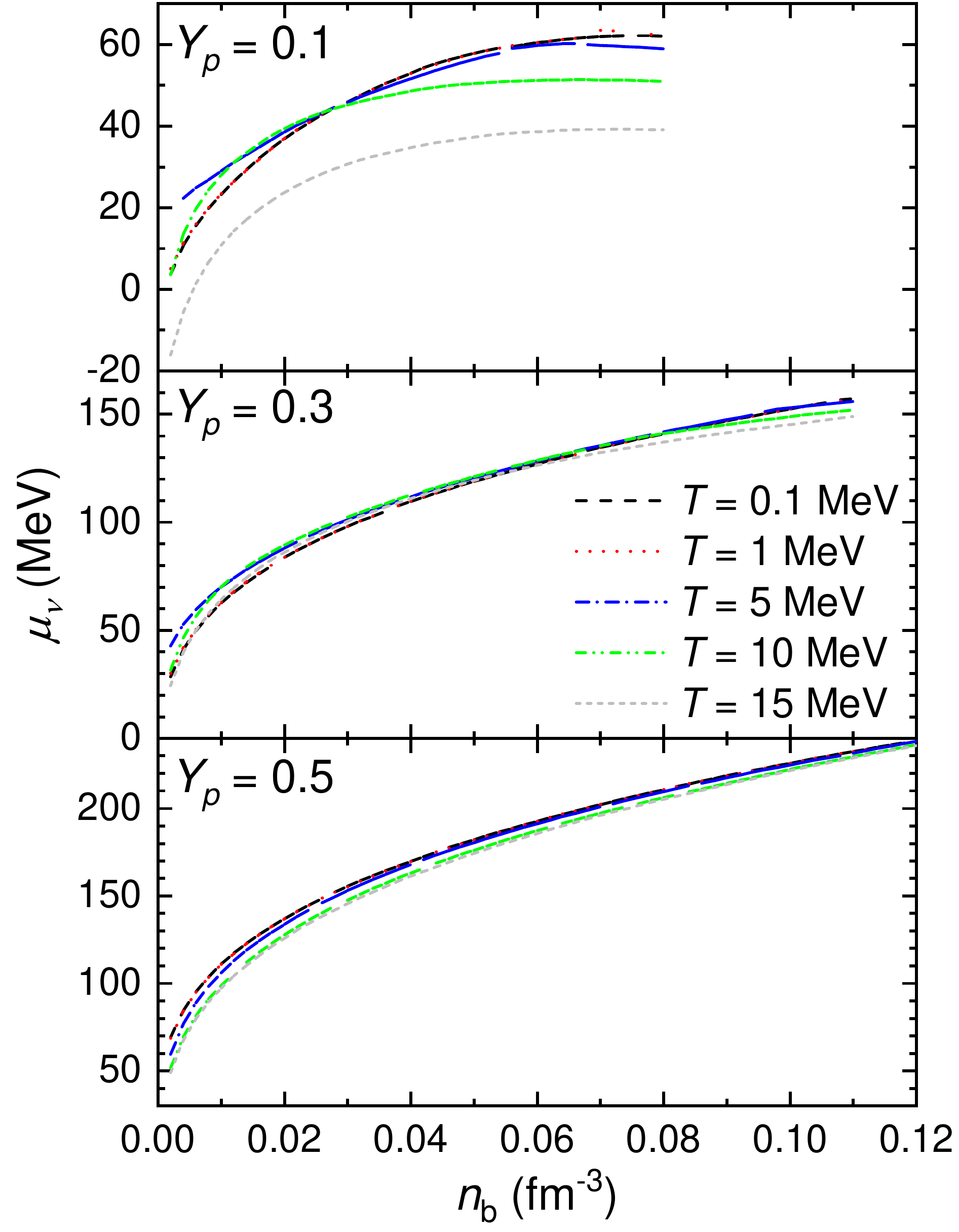}
\caption{\label{Fig:munv} Equilibrium electron-neutrino chemical potential $\mu_{\nu}$ ($=\mu_e+\mu_p-\mu_n$) for the nuclear matter in correspondence to Fig.~\ref{Fig:FpA}.}
\end{figure}

For realistic astromaterials in supernova,  proto-neutron stars, and binary neutron star mergers, neutrinos are trapped in a short period of time and play important roles. As neutrinos do not participate with the strong and electromagnetic interactions, they are distributed uniformly inside WS cells. The presence of trapped neutrinos delays the conversion of protons into neutrons and keeps a relatively large proton fraction for nuclear matter, e.g., $Y_p\approx 0.3$, which favors the formation of various inhomogeneous structures as indicated in Fig.~\ref{Fig:FpA}. Due to the neutrino contributions to the pressure and energy density~\cite{Maruyama2013}, the equation of state is altered and deviates from those presented in Fig.~\ref{Fig:FpA}. In Fig.~\ref{Fig:munv} we present the equilibrium electron-neutrino chemical potentials as functions of density, which are fixed with $\mu_{\nu}=\mu_e+\mu_p-\mu_n$. For equilibrated systems with trapped neutrinos, the contributions of neutrinos increase with $\mu_{\nu}$, which are increasing with density $n_\mathrm{b}$ and proton fraction $Y_p$. The impact of temperature is less significant, where the equilibrium electron-neutrino chemical potential generally decreases with $T$. Note that if we fix the lepton fraction $Y_l=Y_p+N_\nu/A$, the obtained proton fraction $Y_p$ and electron-neutrino chemical potential $\mu_{\nu}$ will deviate slightly from the values indicated in Fig.~\ref{Fig:munv}. More detailed discussions can be found in earlier publications, e.g., those in Ref.~\cite{Maruyama2013}.

Based on the density profiles of nuclear pasta illustrated in Fig.~\ref{Fig:Dens}, the droplet size $R_\mathrm{d}$ and volume of WS cells $V$ can be obtained with
\begin{eqnarray}
 R_\mathrm{d} &=&
 \left\{\begin{array}{l}
   R_\mathrm{W}\left(\frac{\langle n_p \rangle^2}{\langle n_p^2 \rangle}\right)^{1/D},  \text{\ \ \ \ \ \ \  droplet-like}\\
   R_\mathrm{W} \left(1- \frac{\langle n_p \rangle^2}{\langle n_p^2 \rangle}\right)^{1/D},  \text{\ \ bubble-like}\\
 \end{array}\right.,  \label{Eq:Rd} \\
  V &=&
 \left\{\begin{array}{l}
   \frac{4}{3}\pi R_\mathrm{W}^3,\  D = 3\\
   \pi l R_\mathrm{W}^2 , \  D = 2\\
   l^2 R_\mathrm{W}, \ \  D = 1\\
 \end{array}\right.. \label{Eq:V}
\end{eqnarray}
where $\langle n_p^2 \rangle = \int n_p^2(\vec{r}) \mbox{d}^3 r/V$ and $\langle n_p \rangle  = \int n_p(\vec{r}) \mbox{d}^3 r/V$. The parameter $D$ stands for the dimension, where $D = 3$ corresponds to droplets/bubbles, $D = 2$ to rods/tubes, and $D = 1$ to slabs. Note that for the cases with $D = 1$ and 2, we have introduced an additional cell size $l$ so that the volume $V$ is finite, which is essentially a random number and we take $l=10\sqrt{3}$ fm and 20 fm for $D = 1$ and 2. The proton number of the system is then fixed by
\begin{equation}
  Z=n_\mathrm{b}Y_p V.  \label{Eq:Z}
\end{equation}

\begin{figure*}[htbp]
\begin{minipage}[t]{0.335\linewidth}
\centering
\includegraphics[width=\textwidth]{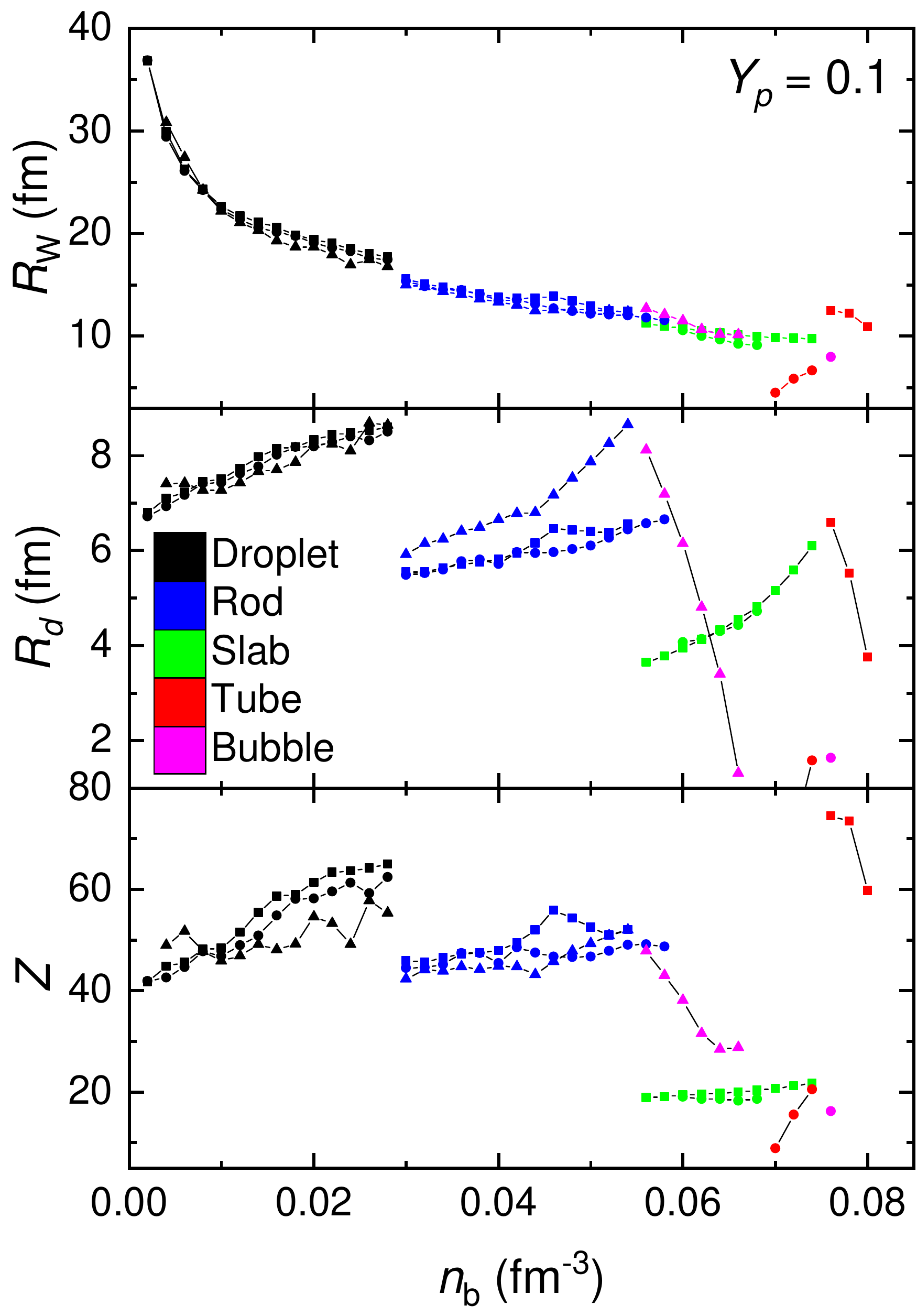}
\end{minipage}%
\hfill
\begin{minipage}[t]{0.318\linewidth}
\centering
\includegraphics[width=\textwidth]{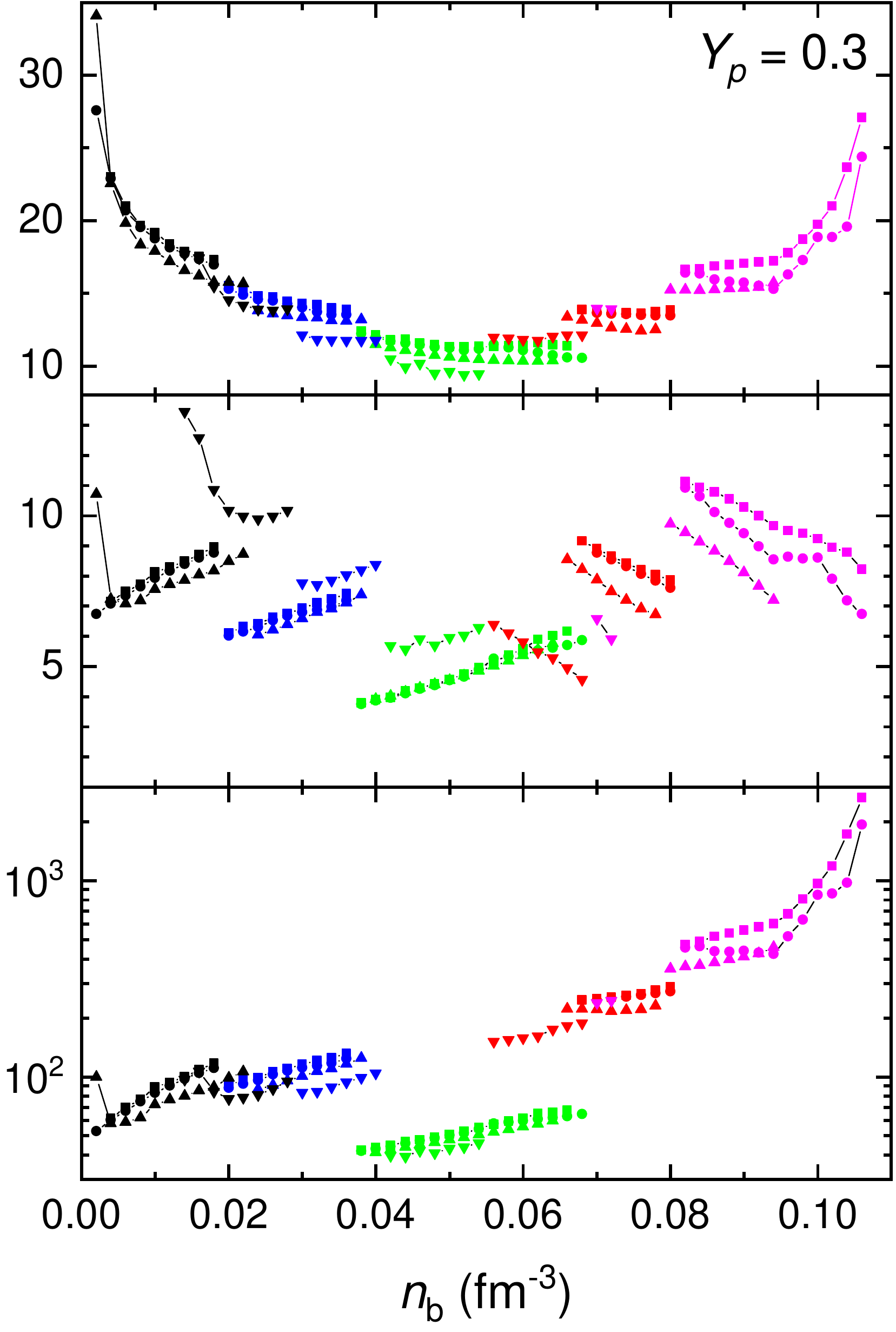}
\end{minipage}
\hfill
\begin{minipage}[t]{0.333\linewidth}
\centering
\includegraphics[width=\textwidth]{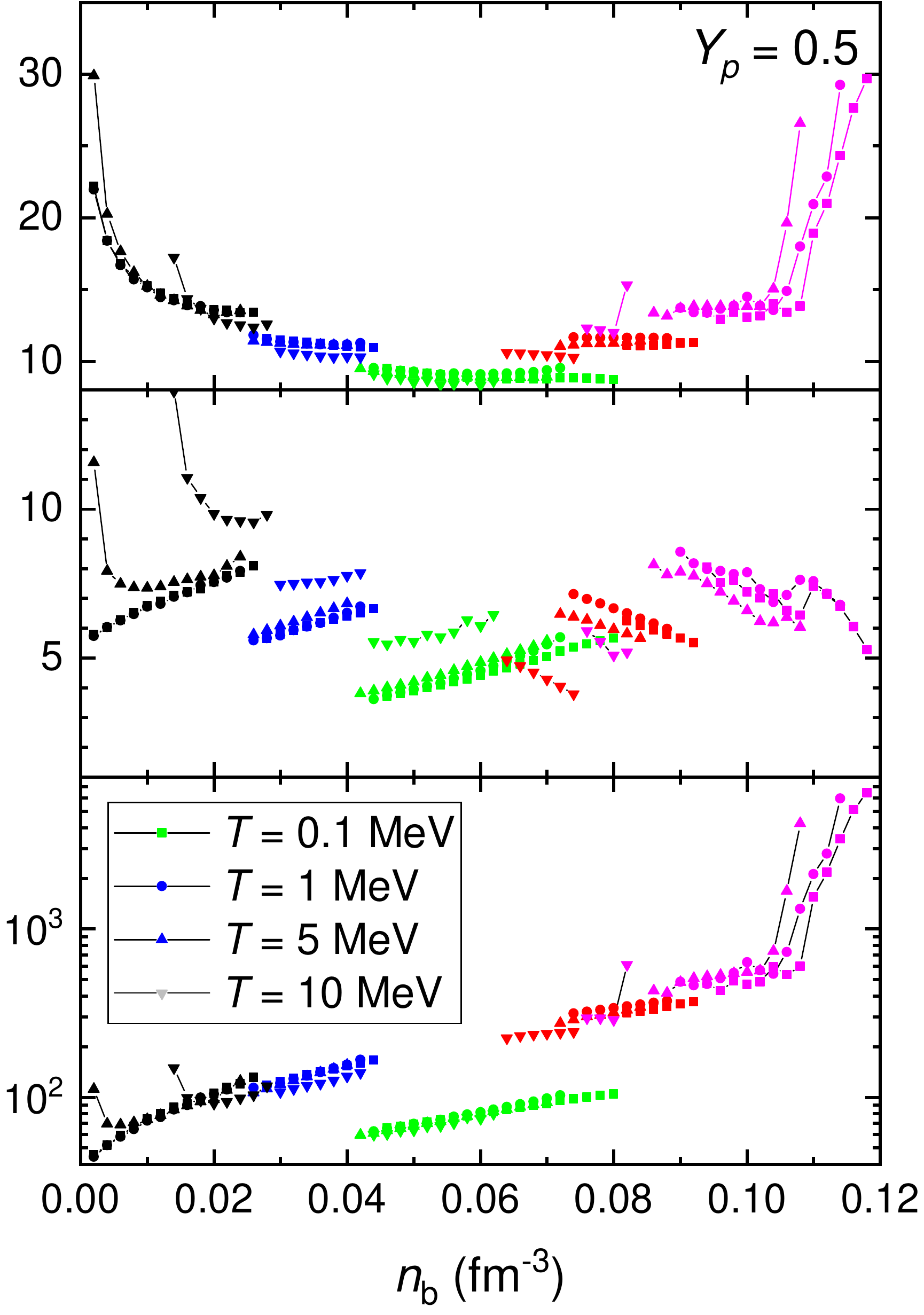}
\end{minipage}
\caption{\label{Fig:Mic} Proton number $Z$, droplet/bubble size $R_\mathrm{d}$, and WS cell size $R_\mathrm{W}$ for the nonuniform nuclear matter in correspondence to Fig.~\ref{Fig:FpA}. }
\end{figure*}

In Fig.~\ref{Fig:Mic} we present the proton number $Z$, droplet/bubble size $R_\mathrm{d}$, and WS cell size $R_\mathrm{W}$ for optimum nuclear pasta structures, which are obtained with Eqs.~(\ref{Eq:Rd}-\ref{Eq:Z}) adopting spherical or cylindrical approximations for WS cells. As illustrated in Fig.~\ref{Fig:Dens}, the microscopic structures of nuclear droplets are altered as we increase $T$, while that of the WS cell size $R_\mathrm{W}$ remains almost constant. Similar situations are observed in Fig.~\ref{Fig:Mic}, where as we increase $T$ the droplet size becomes larger and bubble size smaller. This is mainly because the density distribution inside WS cells approaches to the limit of a uniform one as $T$ increases. Meanwhile, it is found that the WS cell size $R_\mathrm{W}$ and proton number $Z$ decrease slightly with $T$, which would nonetheless increase drastically approaching to the uniform-nonuniform phase boundaries. The reason for the evolutions of $R_\mathrm{W}$ and $Z$ with respect to $T$ is twofold. On the one hand, the reduction of $R_\mathrm{W}$ is mainly attributed to the reduction of surface tension between the liquid and gas phases of nuclear matter, where the corresponding densities become similar as $T$ increases~\cite{Maruyama2010_NPA834-561c}. On the other hand, as one approaches to the uniform-nonuniform phase boundaries, the proton fractions in each phases start to take similar values with the enhancement of the congruence~\cite{Maruyama2011_JPCS312-042015, Maruyama2012_AIPCP1441-387}, where $R_\mathrm{W}$ grows drastically as nuclear pasta resembles the liquid-gas mixed phase obtained with Maxwell construction. Comparing the cases adopting different proton fractions $Y_p$, the optimum WS cell size generally decreases with $Y_p$ in order to reduce the Coulomb energy, which is expected to be proportional to the surface energy according to the compressible-liquid-drop model~\cite{Ravenhall1983_PRL50-2066}. The droplet/bubble sizes vary slightly with $Y_p$, so that the surface areas are almost the same, indicating similar surface energies. Meanwhile, we find that the proton number $Z$ normally increases with $Y_p$. As we increase $n_\mathrm{b}$, it is found that $R_\mathrm{W}$ decreases for the droplet-like phases (droplet, rod, slab), while this trend reverses for the tube and bubble phases. The droplet size $R_\mathrm{d}$ and proton number $Z$ are normally increasing with $n_\mathrm{b}$ aside from the cases close to the uniform-nonuniform phase boundaries, while the bubble size $R_\mathrm{d}$ decreases with $n_\mathrm{b}$.

\subsection{\label{sec:pasta_lat} Lattice structures}

The nonuniform structures of nuclear matter presented in Sec.~\ref{sec:pasta_bulk} are obtained assuming geometrical symmetries for the WS cells, where the interaction among different cells were neglected. For more realistic cases, the droplets, rods, slabs, tubes, and bubbles are expected to form various crystalline structures. For example, the droplets/bubbles could form SC, BCC, and FCC lattices, where at small enough densities the BCC lattice is the most stable configuration~\cite{Oyamatsu1984_PTP72-373}. Nevertheless, as density increases, the FCC lattice may become more favorable~\cite{Okamoto2012_PLB713-284, Okamoto2013_PRC88-025801, Xia2021_PRC103-055812}. In particular, the covariant density functional MTVTC adopted here predicts stable FCC lattice for nuclear droplets in neutron stars at $n_\mathrm{b}=0.06\ \mathrm{fm}^{-3}$~\cite{Xia2021_PRC103-055812}. It is interesting to compare our results with that of Coulomb crystals considering the collective (phonon) degrees of freedom, where the FCC lattice may become more stable than BCC lattice for certain ion charge numbers and densities~\cite{Baiko2002_PRE66-056405}. Since the contribution of phonons is neglected in our study, FCC lattice becomes more tightly bound than BCC lattice due to other contributions such as the dripped neutrons and nonzero sizes of droplets, while more detailed investigations with phonons shall be carried out in our future study. The rods/tubes could form simple and honeycomb configurations, where the honeycomb one were found to be more stable~\cite{Okamoto2012_PLB713-284, Okamoto2013_PRC88-025801, Xia2021_PRC103-055812}. Very complicated structure may be formed by slabs, e.g.,  the primitive, gyroid, diamond morphologies~\cite{Schuetrumpf2020_PRC101-055804}, nuclear waffles~\cite{Schneider2014_PRC90-055805, Sagert2016_PRC93-055801}, and Parking-garage structures~\cite{Berry2016_PRC94-055801}. In such cases, based on the nuclear pasta structures obtained assuming geometrically symmetric WS cells, we further investigate the corresponding crystalline structures in a three-dimensional geometry with reflection symmetry~\cite{Xia2021_PRC103-055812}, which is equivalent to considering only one octant of the unit cell.

\subsubsection{\label{sec:pasta_coext} Coexistence of various configurations}

\begin{figure}
\includegraphics[width=\linewidth]{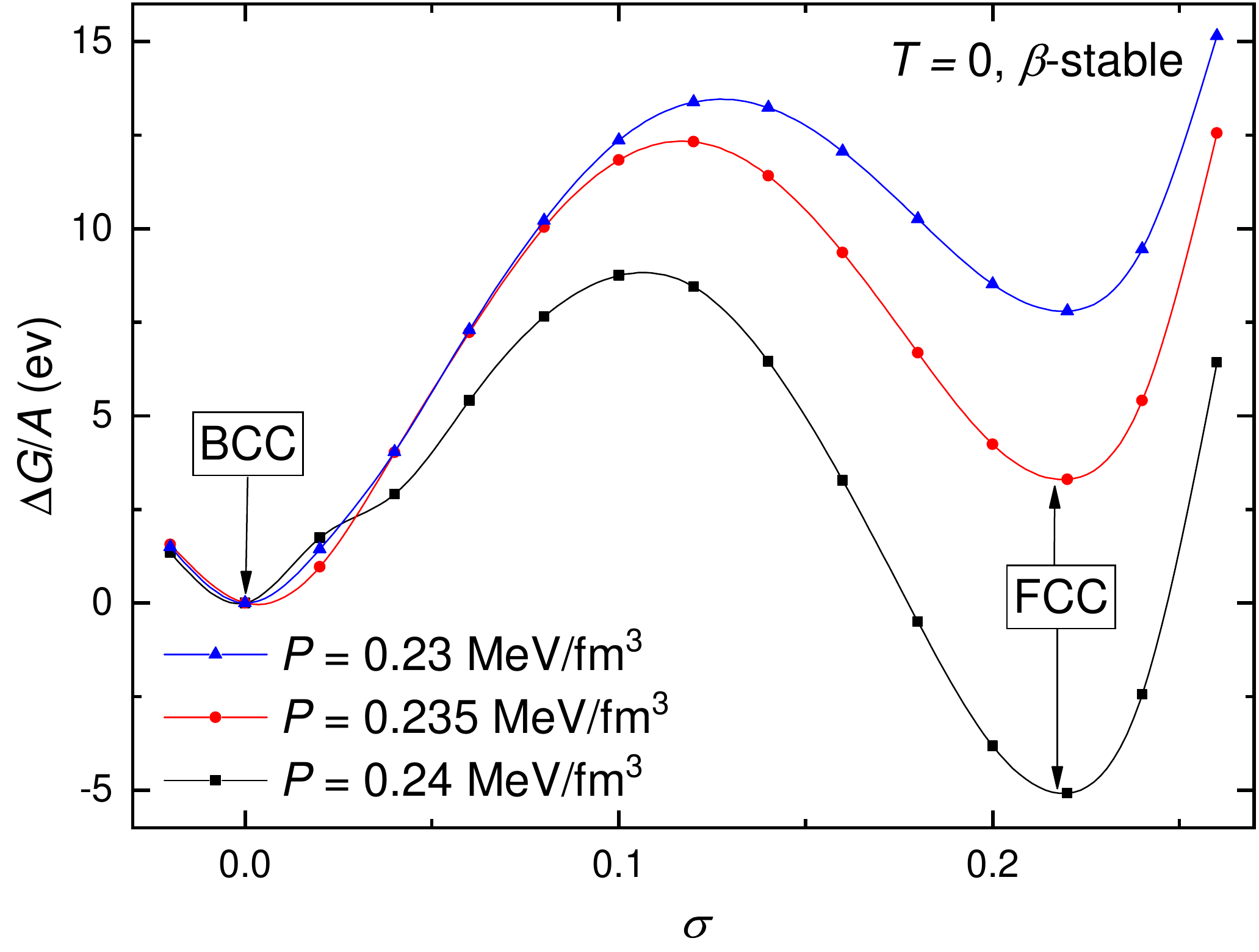}
\caption{\label{Fig:DEBtoF} {Variation of Gibbs free energy per baryon $\Delta G/A$ as a function of $\sigma$, where the droplets in BCC lattice ($\sigma=0$) evolve into FCC lattice ($\sigma=\sigma_\mathrm{F}$) at fixed pressures around $n_\mathrm{b}\approx0.06\ \mathrm{fm}^{-3}$.} The $\beta$-stability condition is always fulfilled.}
\end{figure}

In principle, as temperature increases, not only the internal structures of droplets are modified as indicated in Fig.~\ref{Fig:Dens}, the droplet will gain kinetic energy as well and constantly relocate itself, causing transitions among different lattice structures. For example, the BCC lattice can evolve into FCC lattice by introducing a displacement on $z$-axis with $\sigma=\sigma_\mathrm{F}=2-2^{5/6}\approx 0.2182$, i.e., the Bain path~\cite{Bain1924_TAIMME70-25}, where a droplet at position ($x,y,z$) is moved to a new position ($x',y',z'$) with
\begin{equation}
  z' = z \left( 1-\frac{\sigma}{2} \right) ^{-2},
  x' = x \left( 1-\frac{\sigma}{2} \right),
  y' = y \left( 1-\frac{\sigma}{2} \right). \label{Eq:deform}
\end{equation}
In Fig.~\ref{Fig:DEBtoF} we present the variation of Gibbs free energy per baryon $\Delta G/A$ with respect to the droplet phase in BCC lattice, where the Gibbs free energy is fixed by $G = F - \sum_i \mu_i N_i$ with $N_p=N_e\approx 22$ and $N_n\approx 1442$ for each unit cell. It is evident that droplets in FCC lattice become more favorable than BCC lattice as we increase the pressure, i.e., $\Delta G(\sigma_\mathrm{F})<0$~\cite{Xia2021_PRC103-055812}. Nevertheless, their difference in the Gibbs free energy per baryon is rather small. In such cases, for nuclear matter with finite temperatures, various crystalline structures and nuclear shapes may coexist and form polycrystalline configurations. The barrier height that separates the BCC and FCC configurations should be fixed by multiplying the number of nucleons that move simultaneously as the transition from one phase to another takes place, which is expected to be large. In such cases, at temperatures below the barrier height, the transition between different phases could only occur via quantum tunneling or seismic activities that involve large deformations, which are expected to take place on a much larger timescale. Due to the shell effects of nuclei, in addition to the lattice structures, there may exist extensive local minima in the potential energy surfaces of nuclear shapes, which may still be populated if the temperature drops quickly, leading to the formation of an amorphous solid in neutron star crusts~\cite{Newton2022_PRC105-025806}.

\begin{figure*}[htbp]
\begin{minipage}[t]{0.349\linewidth}
\centering
\includegraphics[width=\textwidth]{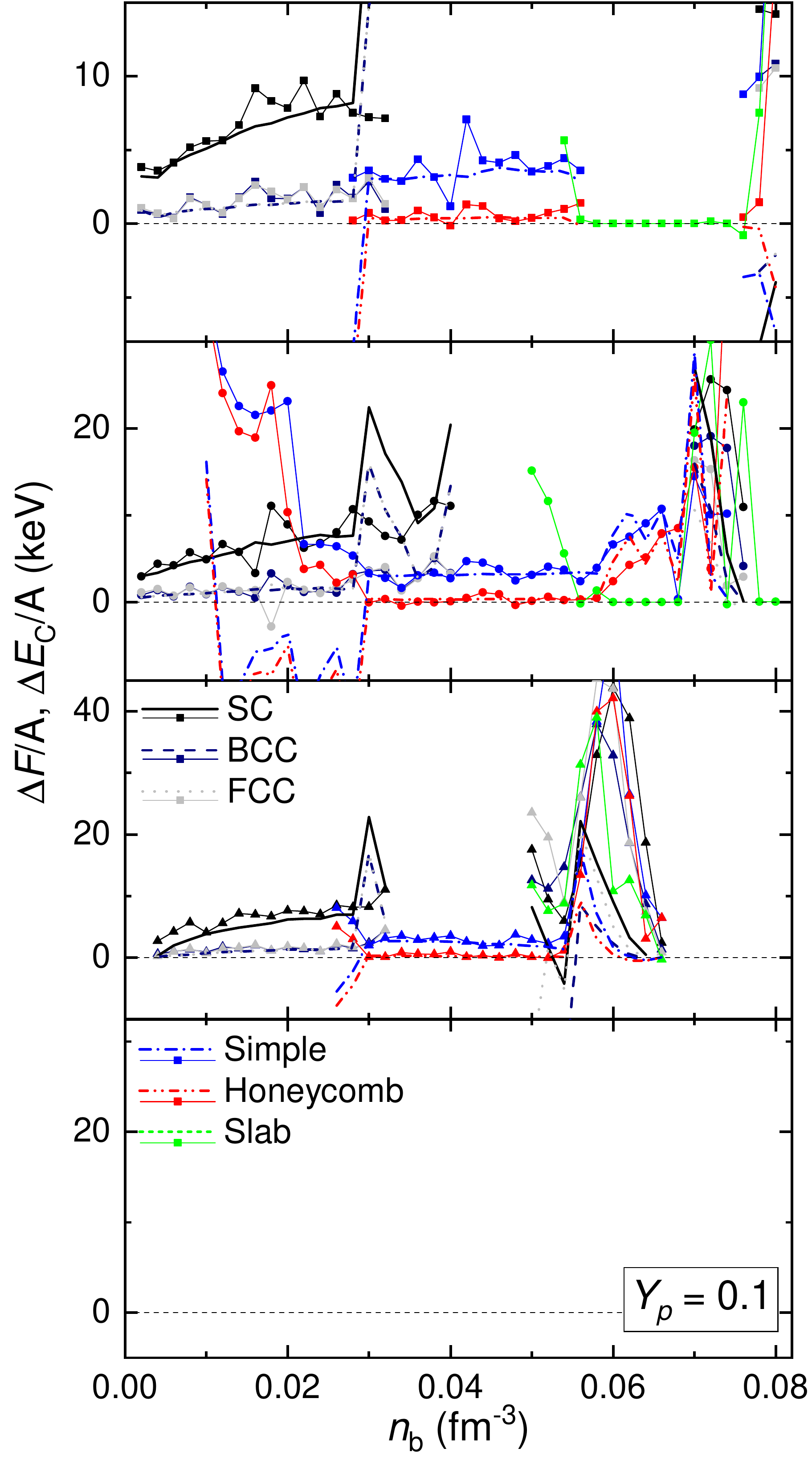}
\end{minipage}%
\hfill
\begin{minipage}[t]{0.311\linewidth}
\centering
\includegraphics[width=\textwidth]{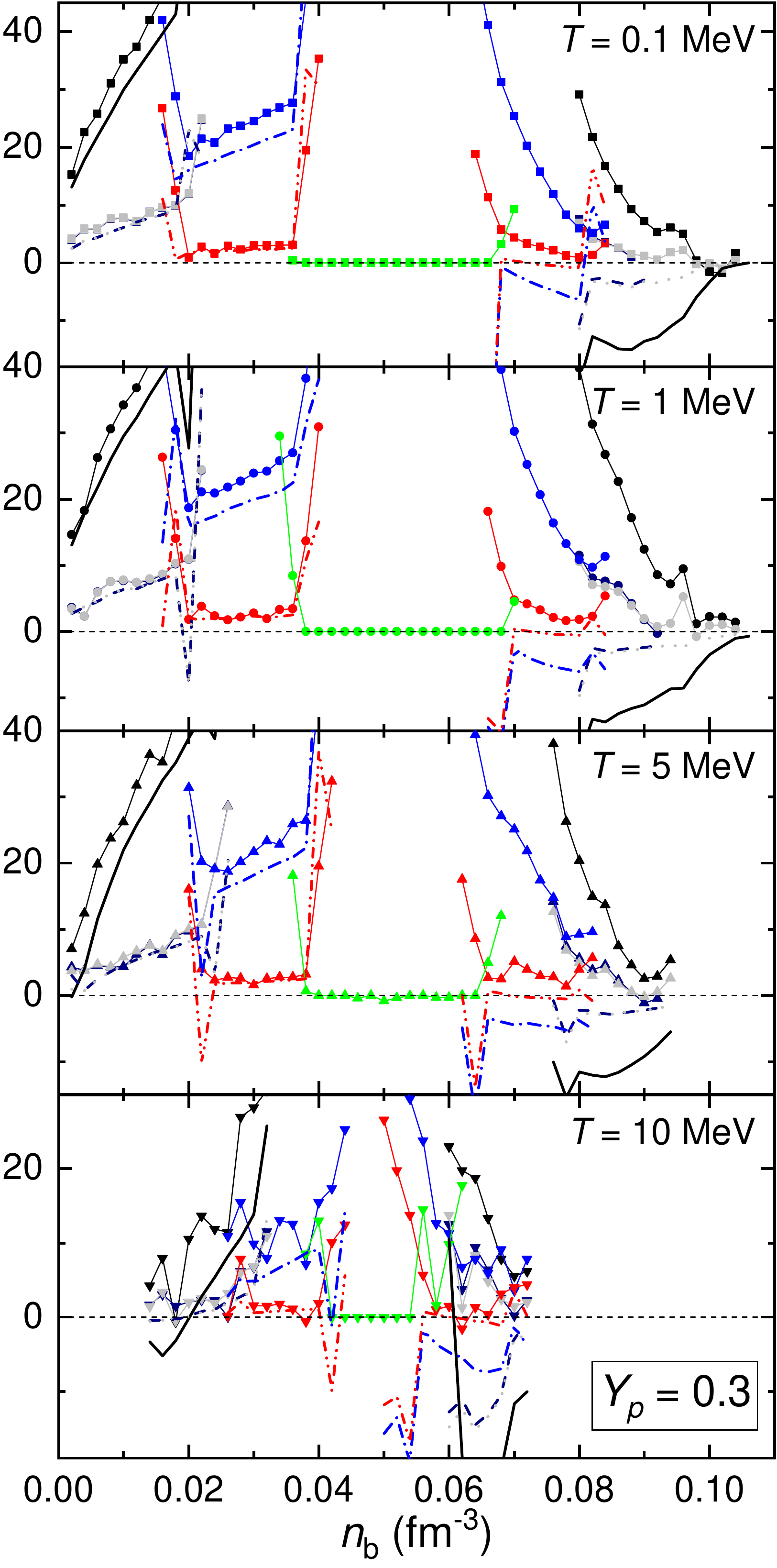}
\end{minipage}
\hfill
\begin{minipage}[t]{0.32\linewidth}
\centering
\includegraphics[width=\textwidth]{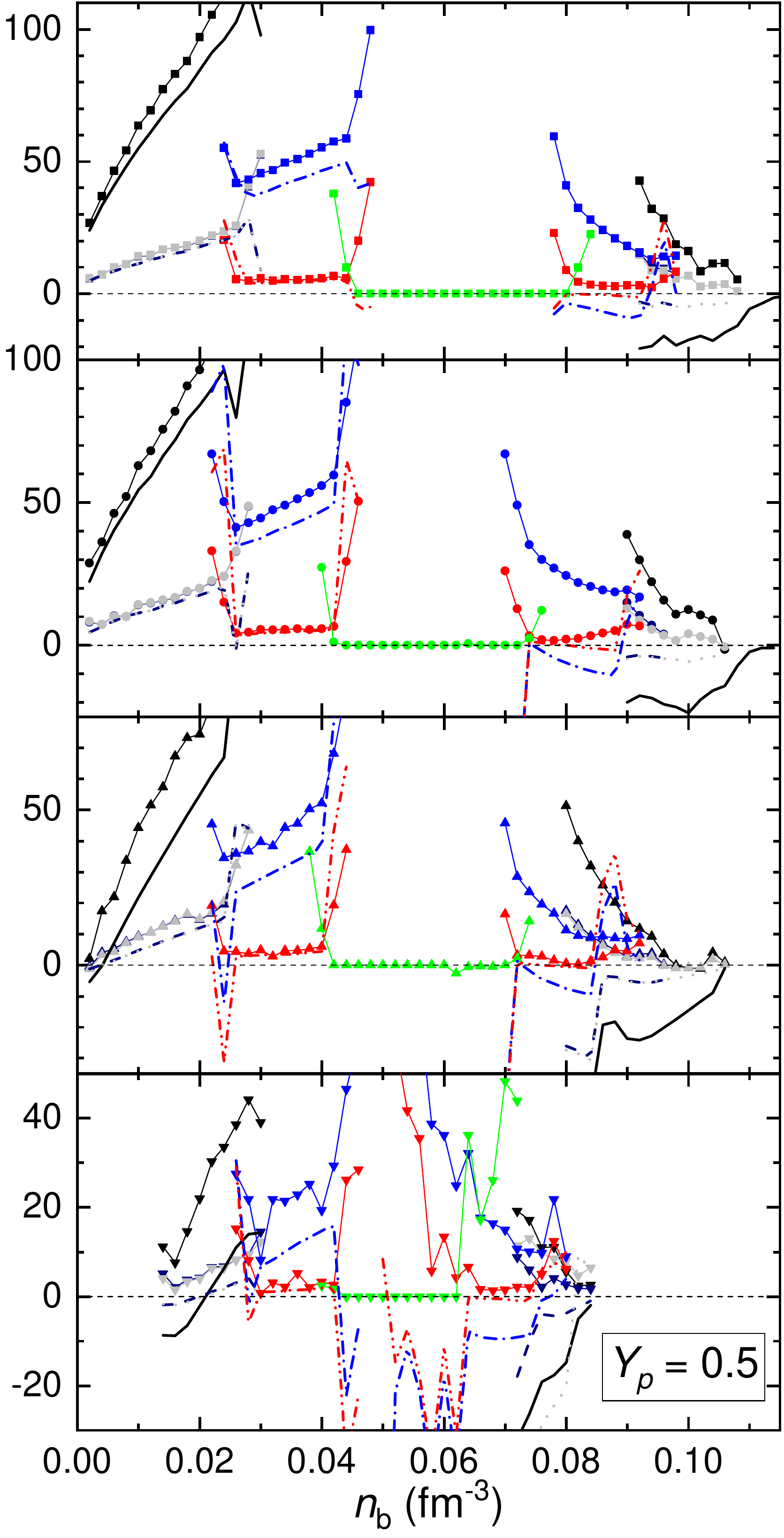}
\end{minipage}
\caption{\label{Fig:DFA} Free energy excess per baryon ($\Delta F/A$, symbols connected by lines) and Coulomb energy excess per baryon ($\Delta E_\mathrm{C}/A$, lines) for various nuclear pasta configurations with respect to those indicated in Fig.~\ref{Fig:FpA}. The lattice structures are marked with different colors and line styles, while different temperatures and proton fractions are adopted for each panel.}
\end{figure*}

In this work we consider six types of lattice structures for nuclear pasta with the droplets/bubbles forming SC, BCC, and FCC lattices, the rods/tubes forming simple and honeycomb configurations, and slabs. In principle, we should examine all possible lattice structures and nuclear shapes. However, more exotic shapes are expected to have much larger free energies in the absence of shell effects~\cite{Okamoto2012_PLB713-284, Okamoto2013_PRC88-025801, Xia2021_PRC103-055812}, which makes them less important and we thus leave this topic for our future study. In Fig.~\ref{Fig:DFA} we present the free energy excess per baryon and Coulomb energy excess per baryon for various nuclear pasta configurations obtained in a three-dimensional geometry, which are fixed by subtracting the free energies and Coulomb energies of the most stable configurations obtained with spherical or cylindrical approximations for WS cells as indicated in Fig.~\ref{Fig:FpA}. The free energy excesses are thus mainly from the interactions among different cells and vary with lattice structures, i.e., the lattice energies, which decrease with temperature $T$ and increase with proton fraction $Y_p$. This is mainly because nucleons inside unit cells become more uniformly distributed as we increase $T$ or decrease $Y_p$.

\begin{figure}
\centering
\includegraphics[width=\linewidth]{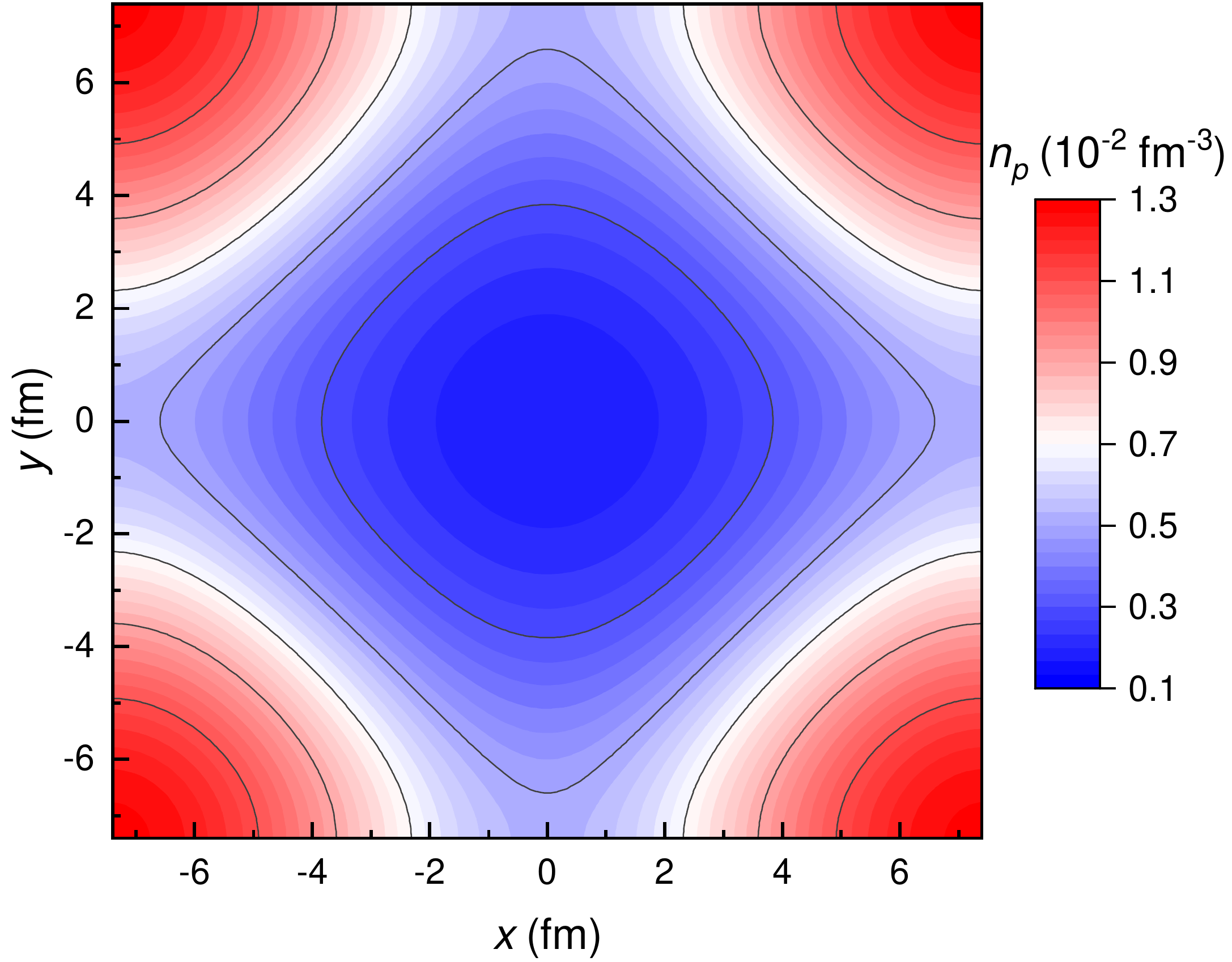}
\caption{\label{Fig:Tube} Proton density profile for the tube phase in simple configuration obtained at $n_\mathrm{b}=0.06$ fm$^{-3}$, $Y_p=0.1$, and $T=5$ MeV.}
\end{figure}

Comparing  $\Delta F/A$ with $\Delta E_\mathrm{C}/A$, it is evident that at small temperatures and densities the free energy excesses are mainly from the Coulomb interaction among cells, i.e., $\Delta F/A\approx \Delta E_\mathrm{C}/A$. However, this relation quickly fails  as we increase $n_\mathrm{b}$ and $T$, where the interaction among nucleons start to play an important role. In particular, at large densities with the emergence of bubble-like structures, $\Delta F/A$ deviates significantly from $\Delta E_\mathrm{C}/A$. This is mainly caused by the additional contributions of nuclear interactions, where nucleons relocate themselves outside of the bubble. As an example, in Fig.~\ref{Fig:Tube} we present the proton density profile for the tube phase in simple configuration obtained at $n_\mathrm{b}=0.06$ fm$^{-3}$, $Y_p=0.1$, and $T=5$ MeV. It is evident that $n_p$ at the boundaries of the unit cell does not follow the cylindrical symmetry, where the protons form clusters on the four corners. In such cases, Coulomb energy $\Delta E_\mathrm{C}/A$ alone does not account for the lattice energy $\Delta F/A$, while the relocation of nucleons and strong interactions among them have to be considered. The difference between $\Delta E_\mathrm{C}/A$ and $\Delta F/A$ decrease as the density approaches to the uniform-nonuniform phase boundaries with $\Delta F/A\rightarrow 0$ and $\Delta E_\mathrm{C}/A \rightarrow 0$.

Similar to our previous findings, for the droplet/bubble phases, the free energy per baryon of BCC and FCC lattices are almost indistinguishable with BCC lattice being slightly more stable at small densities~\cite{Oyamatsu1984_PTP72-373}, while the SC lattices are typically unstable in comparison with BCC/FCC lattices except for few cases in the bubble phases. For the rod/tube phases, the honeycomb configuration is always more stable than the simple one. Note that the numerical uncertainty in Fig.~\ref{Fig:DFA} is on the order of $\sim$keV, which is sufficient considering the high temperatures adopted here. The probability of various combinations of nuclear shapes and lattice structures is expected to follow the statistical distribution~\cite{Raduta2019_NPA983-252}
\begin{equation}
  \text{Probability}\propto\exp\left(-\Delta G A_\mathrm{d}/AT\right). \label{Eq:prob}
\end{equation}
Here $A_\mathrm{d}$ represents the number of nucleons that move collectively in nuclear pasta, which is just the baryon number of each droplet at small enough densities and temperatures. For rod/tube and slab phases, as they extend infinitely in space, the value for $A_\mathrm{d}$ is not so straightforward and approaches to infinity. However, as will be illustrated in Sec.~\ref{sec:pasta_therm}, the one-dimensional ordering for infinite rods, tubes, and slabs are expected to be destroyed by thermodynamic fluctuations, effectively making the collective nucleon number $A_\mathrm{d}$ finite. The value for $A_\mathrm{d}$ is expected to decrease with temperature and finally reaches 1 once nuclear matter becomes uniform, where the degrees of freedom for all nucleons are effectively released.

\subsubsection{\label{sec:pasta_therm} Thermodynamic fluctuations}

\begin{figure*}[htbp]
\includegraphics[width=0.7\linewidth]{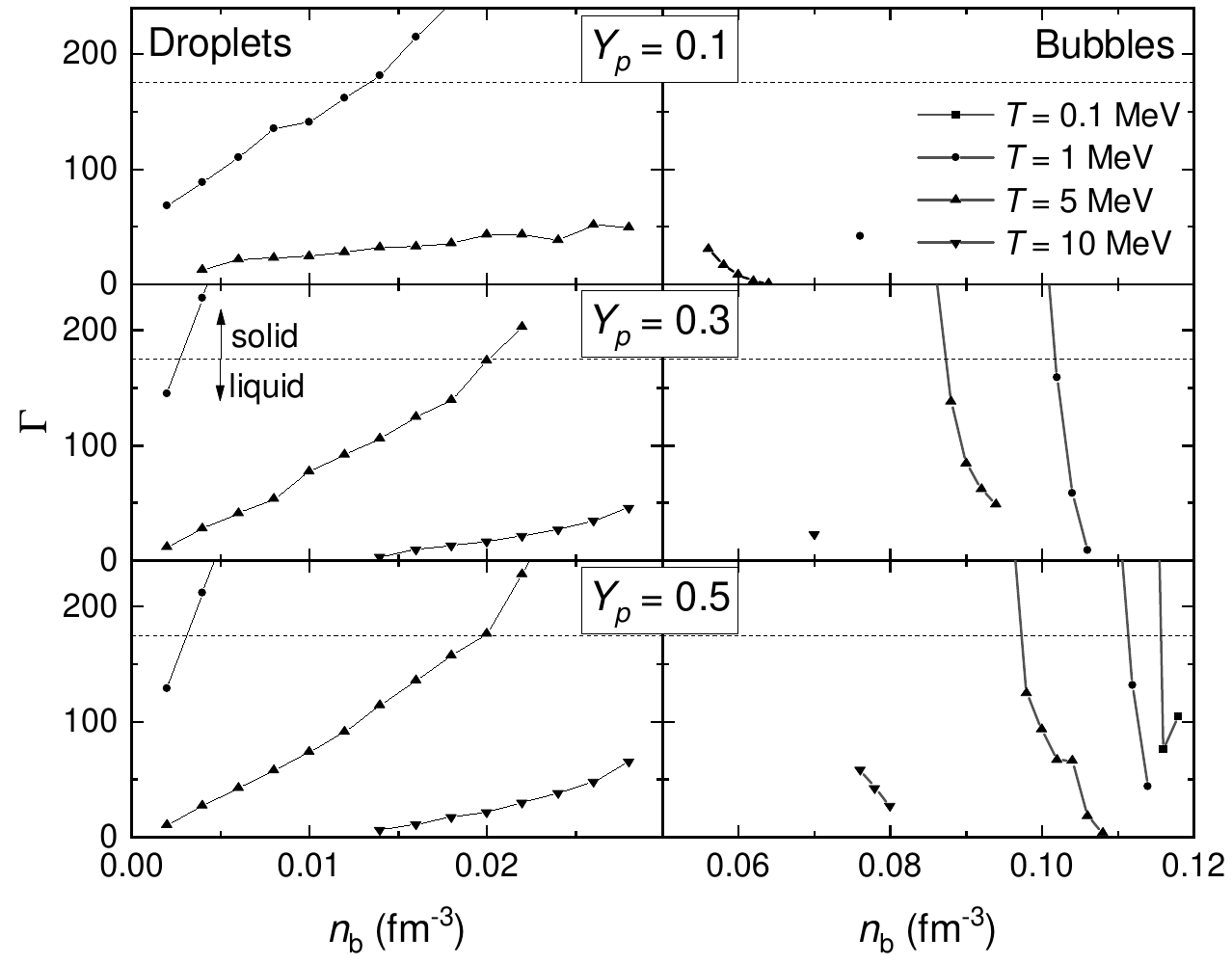}
\caption{\label{Fig:Gamma} Coulomb parameter of nuclear droplets (Left) and bubbles (Right) obtained with Eq.~(\ref{Eq:Gamma}), which corresponds to those indicated in Fig.~\ref{Fig:Mic}. The horizontal lines mark the melting value for $\Gamma$, above which those nuclear droplets/bubbles freeze and form stable crystalline structures~\cite{Brush1966_JCP45-2102, Potekhin2000_PRE62-8554}.}
\end{figure*}

If we increase temperature, the thermodynamic fluctuations of droplets become significant, which eventually lead to melting and form a liquid of droplets. Considering only Coulomb interaction among droplets with charge screening of electrons, the melting temperature of one-component plasma can be estimated with the Coulomb parameter
\begin{equation}
  \Gamma=\frac{Z^2e^2}{R_\mathrm{W} T}, \label{Eq:Gamma}
\end{equation}
which characterizes the ratio of a typical Coulomb energy to the thermal energy. According to the Monte Carlo simulations, the plasma freezes at $\Gamma\gtrsim 175$~\cite{Brush1966_JCP45-2102, Potekhin2000_PRE62-8554}. Note that the exact criterion may be altered by electron screening~\cite{Potekhin2000_PRE62-8554}. Meanwhile, as indicated in Fig.~\ref{Fig:Dens}, the proton density outside the droplet is not always negligible. In such cases, in order to use the criterion $\Gamma\gtrsim 175$ obtained assuming point-like charges for nuclei, we need to subtract the contribution from the background proton density by replacing $Z$ with $Z_\mathrm{droplet}=Z-n_p(R_\mathrm{W})V$, which roughly accounts for the net charges of nuclei. In the left panels of Fig.~\ref{Fig:Gamma} we present the Coulomb parameter of the nuclear droplets indicated in Fig.~\ref{Fig:Mic}. Evidently, the melting temperatures of the crystalline structures do not align with the uniform-nonuniform transition temperatures derived from Fig.~\ref{Fig:Mic}. At regions beneath the dashed lines, instead of forming crystalline structures, a liquid of droplets takes place. Those droplets will be destabilized further if we increase $T$, where the density profiles become uniform as illustrated in Fig.~\ref{Fig:Dens}. Note that the exact melting temperature may be altered if additional interactions among nucleons are considered, e.g., the additional lattice energy contribution at large $T$ and $n_\mathrm{b}$ as indicated in Fig.~\ref{Fig:DFA}, which can not be accounted for with Coulomb interaction alone.

The effects of temperature is clearly illustrated in Fig.~\ref{Fig:Gamma}, where the Coulomb parameter $\Gamma$ decreases rapidly with $T$. For the droplet phases obtained at $T=0.1$ MeV, the Coulomb parameter $\Gamma\gtrsim 1000$ and increases quickly with density, where the droplets form crystalline structures. As temperature increases, the crystalline structures start to melt, especially for those with small $n_\mathrm{b}$. We note $\Gamma$ increases with proton fraction $Y_p$, which is similar to the cases of uniform-nonuniform phase transitions. Meanwhile, it is found that $\Gamma$ increases with average baryon number density $n_\mathrm{b}$, which is mainly due to the increment of the proton number $Z$ in droplets. The Coulomb parameter of bubble phases can be estimated by replacing the proton number $Z$ in Eq.~(\ref{Eq:Gamma}) with the effective charge number $Z_\mathrm{bubble}=n_p(R_\mathrm{W})V-Z$~\cite{Watanabe2003_PRC68-045801}, where the corresponding values are presented in the right panels of Fig.~\ref{Fig:Gamma}. Similar trends with respect to $T$ and $Y_p$ are observed for the bubble phases. However, in contrast to the droplet phases, we note $\Gamma$ decreases with $n_\mathrm{b}$, which is mainly attributed to the shrinkage of bubbles sizes as indicated in Fig.~\ref{Fig:Mic}.

For the rod, slab, and tube phases, as their WS cells extends infinitely in space, the corresponding proton number $Z$ would become infinitely large. In such cases, it is not likely that those deformed nuclei would vibrate as a whole. Nevertheless, the temperature effects are expected to cause local thermodynamic fluctuations on the shapes of those objects, leading to the destruction of the one-dimensional ordering in an infinite three dimensional system in the context of Landau-Peierls instabilities~\cite{Watanabe2000_NPA676-455}. It was shown that the temperature effects cause defects and nonparallel configurations in slab phases, while weak sinusoidal or hyperbolic splay with a length scale of order the box width was observed as well~\cite{Caplan2021_PRC103-055810}. To roughly estimate disruption caused by thermal fluctuations, we adopt the formalism derived from the liquid-drop model~\cite{Watanabe2000_NPA676-455}, where the mean-square displacements for the slab and rod/tube phases are determined by
\begin{eqnarray}
\langle |\nu|^2 \rangle &\approx& \frac {\sqrt{5} T}{8\pi \varepsilon_\mathrm{C} R_\mathrm{W} \sqrt {1+2u-2u^2}} \ln\left(\frac{L}{a}\right),  \label{Eq:dis_1D}\\
\langle |\nu|^2 \rangle &\approx& \frac{T}{(B+2 C)^{3/4} \sqrt{\pi a\sqrt{2K_3}}}.  \label{Eq:dis_2D}
\end{eqnarray}
with the coefficients
\begin{equation}
B = \frac{3}{2} \varepsilon_\mathrm{C},  \  C \approx 10^{2.1(u-0.3)}\varepsilon_\mathrm{C}, \   K_3 \approx  0.0655 \varepsilon_\mathrm{C} R_\mathrm{W}^2.
\end{equation}
Here $u = \left({R_\mathrm{d}}/{R_\mathrm{W}}\right)^D$ is the volume fraction, $L$ the length scale of slabs, and $\varepsilon_\mathrm{C} = {E_\mathrm{C}}/{V}$ the equilibrium Coulomb energy density with $E_\mathrm{C}=\int (\nabla A)^2 \mbox{d}^3 r/2$, where the interaction among other unit cells are considered. The lattice constant $a$ is obtained with Eqs.~(\ref{Eq:a_2D}) and (\ref{Eq:a_1D}).

\begin{figure}
\centering
\includegraphics[width=\linewidth]{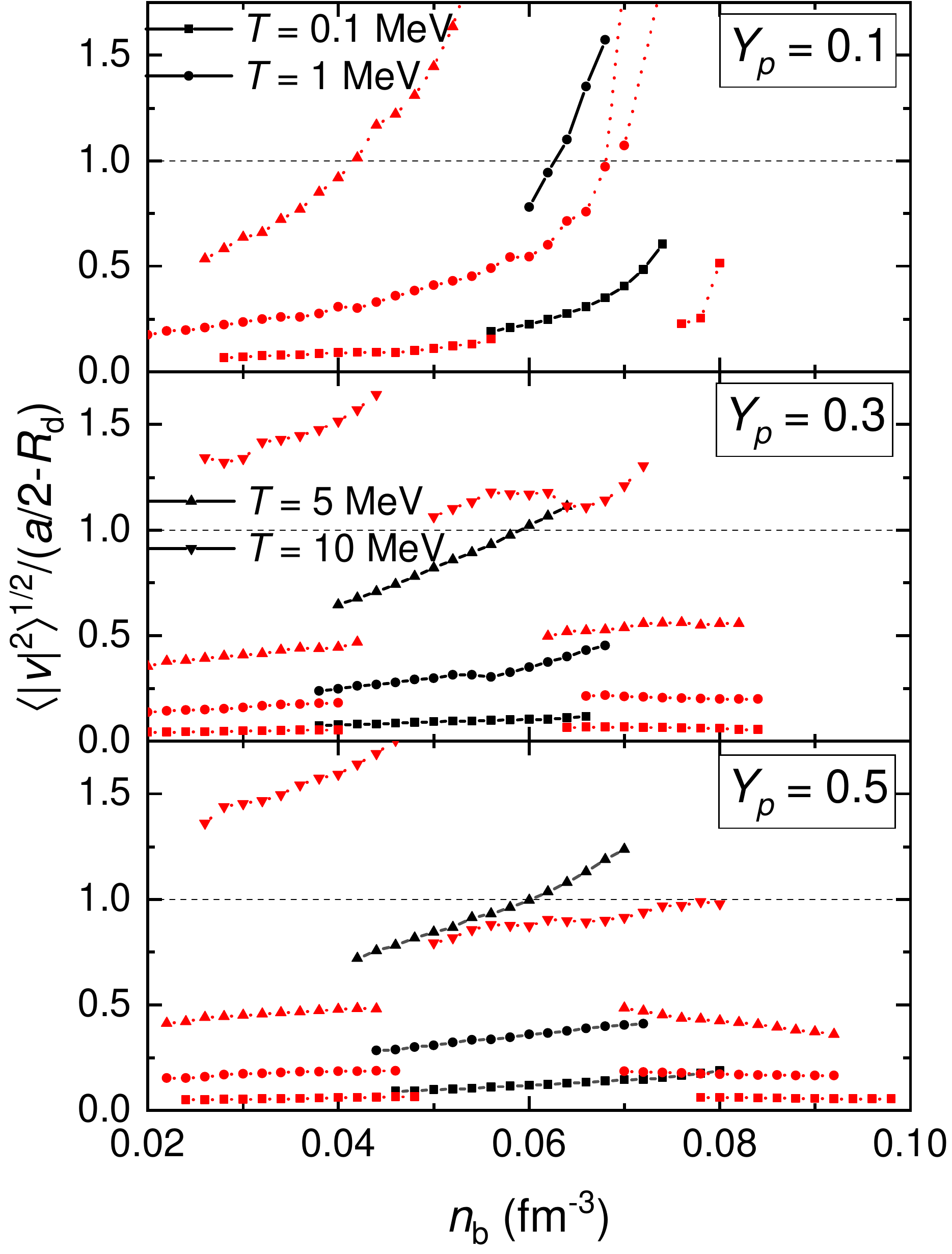}
\caption{\label{Fig:Displace} Root-mean-square displacements of slabs (black-solid curves) and rods/tubes (red-dotted curves) in honeycomb configurations, which are divided by
the distance from the surface of a nucleus to the cell boundary.}
\end{figure}

In Fig.~\ref{Fig:Displace} we present the root-mean-square displacement of slabs and rods/tubes divided by the distance from the surface of a nucleus to the cell boundary $\sqrt{\langle |\nu|^2 \rangle}/(a/2- R_\mathrm{d})$, which are estimated with Eqs.~(\ref{Eq:dis_1D}) and (\ref{Eq:dis_2D}) by taking $L=1\ \mu$m, respectively. If the displacement becomes comparable to the cell size, i.e., $\sqrt{\langle |\nu|^2 \rangle}/(a/2- R_\mathrm{d})\gtrsim 1$, we deem the matter is completely disordered, which is indicated in the region above the horizontal lines. For the rod/tube phases, both the simple and honeycomb configurations are examined in a three-dimensional geometry with the lattice constants fixed by Eq.~(\ref{Eq:a_2D}), where the honeycomb configuration is found to be more stable. In such cases, the displacements indicated with the red-dotted curves correspond to the honeycomb configurations, where the Coulomb energy density $\varepsilon_\mathrm{C}$ from rods/tubes and interaction among cells are derived. Similar to the findings in Ref.~\cite{Watanabe2000_NPA676-455}, the displacement of slabs are usually larger than those of rods/tubes, so that slabs are easily disrupted and form complicated structures~\cite{Caplan2021_PRC103-055810}. The relative displacements generally increase with density $n_\mathrm{b}$ and decrease with proton fraction $Y_p$. It is found that the rods/tubes for nuclear matter at $T=0.1$, 1, 5 MeV and $Y_p=0.3$, 0.5 are generally stable with small displacement, while the slab phases at $T=0.1$, 1 MeV and $Y_p=0.3$, 0.5 are stable. If the proton fraction $Y_p=0.1$ is adopted,  as density increases, both rods and slabs are disrupted for nuclear matter at $T\geq 1$ MeV.

\subsection{Phase diagram}

\begin{figure}
\includegraphics[width=\linewidth]{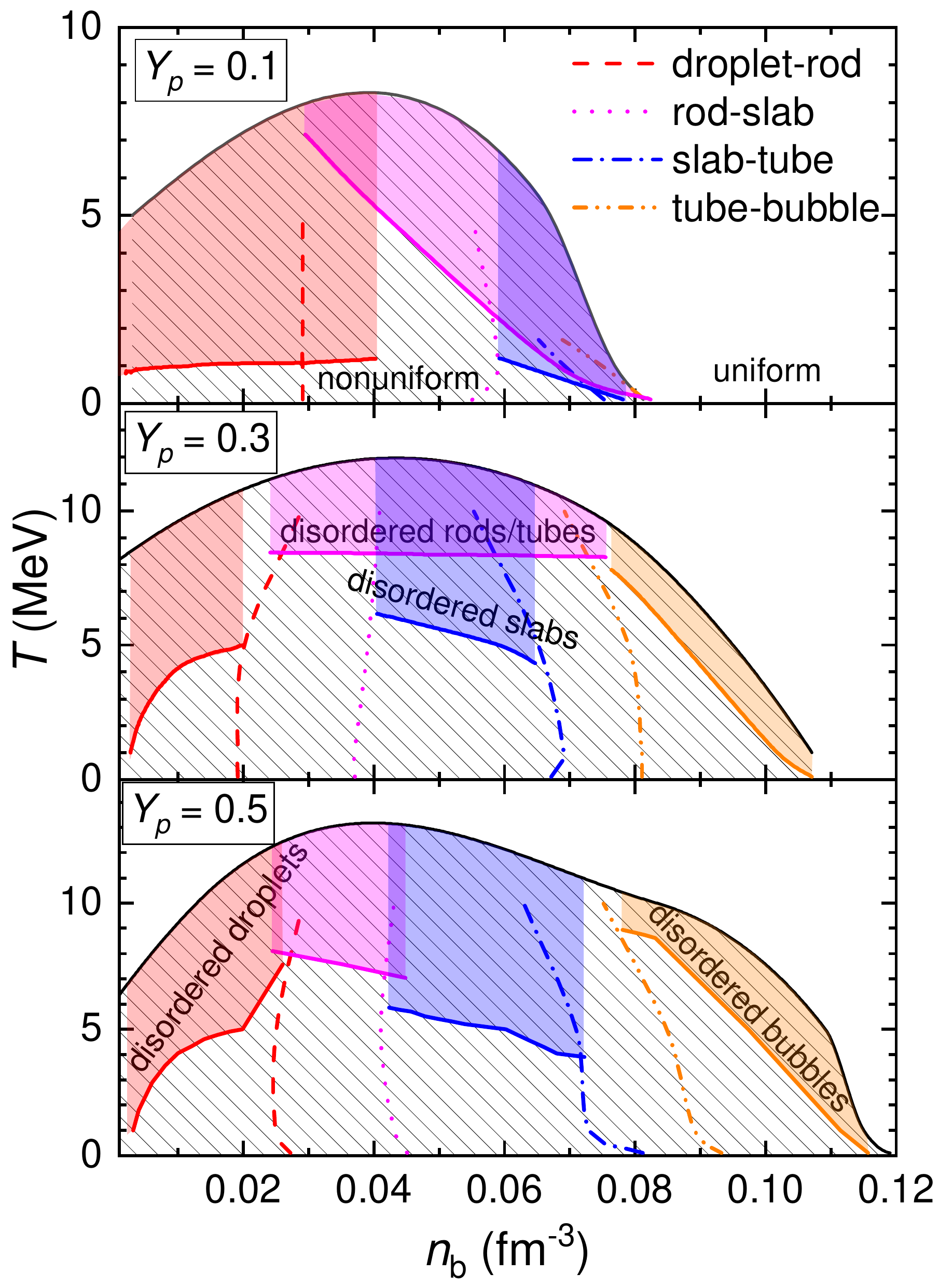}
\caption{\label{Fig:Phase} Rough estimate on the phase diagrams of nuclear matter, where the black-solid curves indicate the uniform-nonuniform phase boundaries. }
\end{figure}

Based on the results presented in Sec.~\ref{sec:pasta_bulk} and Sec.~\ref{sec:pasta_lat}, the phase diagrams of nuclear matter can then be estimated. In Fig.~\ref{Fig:Phase} we present the phase diagrams of nuclear matter at various average densities $n_\mathrm{b}$, temperatures $T$, and proton fractions $Y_p$. The black-solid curves mark the uniform-nonuniform phase boundaries according to the results presented in Fig.~\ref{Fig:FpA} adopting the spherical or cylindrical approximations for WS cells, where the nonuniform phases are indicated by the shaded regions. The colored-solid curves indicate the critical temperatures for the disruption of crystalline structures, which are fixed at $\Gamma=175$ for droplets/bubbles and $\sqrt{\langle |\nu|^2 \rangle}=(a/2- R_\mathrm{d})$ for nonspherical nuclei. The colored regions above those curves mark the temperatures and densities of disordered pasta phases. As we decrease the temperature of uniform nuclear matter and cross the uniform-nonuniform boundary, the density profiles of nuclear matter become nonuniform and form nuclei in various shapes. Nevertheless, those nuclei do not form stable lattice structures or any long-range ordering if the temperatures are above the order-disorder boundaries, which are disordered due to thermodynamic fluctuations and act like liquid. Nuclei in various shapes start to stabilize and freeze into crystalline structures only if the temperatures are below the order-disorder boundaries. The dotted lines indicate the phase boundaries between nuclear shapes, which are fixed by comparing their free energies as indicated in Fig.~\ref{Fig:DFA}. However, the phase boundaries are not strict since various nuclear shapes and crystalline structures could in principle coexist, which follow the statistical probability distribution in Eq.~(\ref{Eq:prob}).

Generally speaking, nuclear pasta structures are destabilized as we decrease the proton fraction, where the critical temperatures for the uniform-nonuniform and order-disorder transitions decrease. The density ranges for nonuniform nuclear matter as well as the nonspherical nuclei decrease as well for smaller $Y_p$. Since the typical proton fraction for nuclear pasta in neutron star crusts is $Y_p\approx 0.02$, the critical temperatures for the phase boundaries become even smaller and reaches $T_\mathrm{c}=0.1$-0.3 MeV at $\Gamma=175$. In such cases, the crystalline structures of nuclear pasta in neutron star crusts are easily disrupted according to the typical proto-neutron star temperatures. The existence of various lattice structures and nuclei shapes could in principle lead to the formation of an amorphous solid if neutron star cools down rapidly~\cite{Newton2022_PRC105-025806}. The phase transition boundaries indicated in Fig.~\ref{Fig:Phase} are expected to affect the properties and evolutions of neutron stars, where the electrical conductivity~\cite{Schneider2016_PRC93-065806}, thermal conductivity~\cite{Horowitz2015_PRL114-031102}, neutrino opacity~\cite{Horowitz2004_PRC69-045804, Schuetrumpf2020_PRC101-055804}, as well as the elastic properties~\cite{Ogata1990_PRA42-4867, Ushomirsky2000_MNRAS319-902, Abbott2020_ApJ902-L21} of neutron star matter are altered.

\section{\label{sec:con}Conclusion}
In this work we have investigated the nuclear pasta structures at high temperatures, where the RMF model with Thomas-Fermi approximation was adopted. The properties and microscopic structures of nuclear pasta were examined adopting spherical or cylindrical approximations for WS cells, where the optimum configurations such as shapes and cell sizes were fixed by minimizing the free energy at fixed temperature $T$ and density $n_\mathrm{b}$. As $T$ increases, the nonuniform structures of nuclear pasta are destabilized, where the density profiles eventually become uniform at large enough $T$. Similar trends are observed as well with respect to the proton fraction $Y_p$, where the nonuniform structures and nonspherical nuclei become less stable at smaller $Y_p$. Consequently, the density range for nonuniform structures of nuclear matter decrease as we increase $T$ or decrease $Y_p$. For equilibrated systems with trapped neutrinos, the contributions of neutrinos increase with density $n_\mathrm{b}$ and proton fraction $Y_p$, while the impact of temperature is less significant. In comparison with the cases of small $T$, the WS cell sizes $R_\mathrm{W}$ and proton numbers $Z$ decrease slightly with $T$ as the surface tension becomes smaller. A rapid growth in $R_\mathrm{W}$ and $Z$ was observed right before the transition to uniform nuclear matter takes place, which is attributed to the enhancement of the congruence with similar proton fractions in the liquid and gas phases of nuclear matter~\cite{Maruyama2011_JPCS312-042015, Maruyama2012_AIPCP1441-387}.

The properties of the nuclear pasta forming various lattice structures were examined in a three-dimensional geometry. Since the equilibrium volume occupied by each nucleus is dominated by its surface and Coulomb energies while the lattice energy is relatively small, varying the lattice structures has little impact on the optimum volume of each nucleus~\cite{Xia2021_PRC103-055812}. In such cases, we investigate nuclear pasta in various lattice structures keeping the volume of each droplet/bubble constant, which is fixed by adopting the volume of a WS cell in its optimum size with the lattice constant $a$ determined by Eqs.~(\ref{Eq:a_3D}-\ref{Eq:a_1D}). At small density and temperatures, the lattice energy is dominated by Coulomb interaction, while the strong interaction among nucleons becomes important at larger $n_\mathrm{b}$ and/or $T$. In particular, at large densities with the emergence of bubble-like structures, nucleons relocate themselves outside of the bubble and form clusters, which deviate from those obtained in cylindrical or spherical symmetric WS cells and alter the interaction among different cells. The differences for the free energies per baryon of nuclear pasta in various shapes and lattice structures are typically on the order of tens of keV. In such cases, different nuclear pasta structures are expected to coexist for nonzero temperatures.

The thermodynamic fluctuations are expected to disrupt the long-range ordering in nuclear pasta structures. We have estimated the the Coulomb parameters $\Gamma$ for the droplet/bubble phases, where the crystalline structures are expected to be destroyed and melt at $\Gamma\lesssim 175$. Similar to the uniform-nonuniform transitions, the crystalline structures are destabilized as we increase $T$ or decrease $Y_p$. As density increases, it was found that $\Gamma$ generally increases for the droplet phase and decreases for the bubble phase. The mean-square displacements in the rods, slabs, and tubes from thermodynamic fluctuations were estimated adopting the formalism derived from the liquid-drop model~\cite{Watanabe2000_NPA676-455}, which generally increase with density. We found that the displacements are increasing with $T$ and decreasing with $Y_p$. Once the root-mean-square displacements become larger than the distance from the surface of a nucleus to the cell boundary, the rods, slabs, and tubes become disordered and behave like liquid. It was found that the rods and tubes are generally more resilient against thermodynamic fluctuations than slabs, which is consistent with previous estimations~\cite{Watanabe2000_NPA676-455}.

The phase diagrams of nuclear matter with respect to density $n_\mathrm{b}$, temperature $T$, and proton fraction $Y_p$ were obtained, which should be useful for various investigations on the properties and evolutions of neutron stars~\cite{Schneider2016_PRC93-065806, Horowitz2015_PRL114-031102, Horowitz2004_PRC69-045804, Schuetrumpf2020_PRC101-055804, Ogata1990_PRA42-4867, Ushomirsky2000_MNRAS319-902, Abbott2020_ApJ902-L21},  supernova dynamics~\cite{Bethe1990_RMP62-801, Watanabe2005_PRL94-031101, Alloy2011_PRC83-035803, Roggero2018_PRC97-045804, Janka2012_ARNPS62-407}, and binary neutron star mergers~\cite{Gamba2019_CQG37-025008, Biswas2019_PRD100-044056, Baiotti2019_PPNP109-103714, LI2020, Gittins2020_PRD101-103025}.

\section*{ACKNOWLEDGMENTS}
This work was supported by the National SKA Program of China (Grant No.~2020SKA0120300), National Natural Science Foundation of China (Grant No.~12275234), and JSPS KAKENHI (Grants No.~20K03951 and No.~20H04742).

\newpage

%

\end{document}